# Numerical Study of Trailing and Leading Vortex Dynamics in a Forced Jet with Coflow

Bharat Bhatia[a] and Ashoke De[a]

[a]Department of Aerospace Engineering, Indian Institute of Technology Kanpur,208016, India.

The interaction of trailing and leading vortex structures in low Reynolds number forced (varicose) jet with coflow is studied using Direct Numerical Simulation (DNS), where we report the dynamics of these vortices for a range of varying parameters. In a circular forced jet with Strouhal number 0.2, continuous formation of these vortices is observed which undergo pairing, tearing and disintegration. The forced jet is analyzed for varying frequency of forced perturbations, coflow temperature, turbulence intensity, momentum thickness, coflow intensity and Reynolds number at a constant forcing amplitude. The observation reveals that any reduction in momentum thickness increases the circulation as well as energy capacity of the leading and the trailing vortices, thereby enhancing the vortex pairing time. However, the hot coflow increases the strength of the leading vortex to such an extent that the generated trailing vortex is weak and dissipates even before it undergoes vortex pairing, while an incomplete disintegration of the leading vortex occurs in this case. Moreover, the strength of leading and trailing vortices is found to decrease with an increase in coflow power, thereby leading to a weaker vortex pairing. The decrease in Reynolds number also reduces the intensity of trailing vortices; whereas, at a lower Reynolds Number (*Re*) the trailing vortices dissipate soon after they form. The findings are achieved based on the analysis done by Proper Orthogonal Decomposition (POD), Dynamic Mode Decomposition (DMD), fast Fourier transform (FFT), Q-criterion and vortex tracking method to understand the dynamic interaction and behavior of the vortices.

______________________________

URL:  ashoke@iitk.ac.in (Ashoke De)

# 1. Introduction

In past few decades, there has been a significant number of studies on the forced jets, vortex structures and their interactions including their effect on jet evolution. Majority of these experimental studies primarily focus on the control of jets, turbulence involved, coherent structures and their interactions. Usually, the external perturbations applied at the jet exit allow to have a particular and localized vortex pairing; in turn, facilitate the visualization of these vortices in the presence of background noise. One of the earliest studies done by Becker *et al.*[1] on the varicose instability of a round jet was subjected to acoustic excitation, where the vortex formation and their pairing generated sub-harmonic frequency at different Reynolds number. Crow and Champaign[2] showed that the preferred mode of a jet was at *Strouhal number (St)= 0.3* where no harmonics were present but reported a generation of maximum turbulence intensity. Later this definition of 'preferred mode' was redefined by Zaman and Hussain[3] based on the maximum fundamental r.m.s. amplitude but not on the turbulence intensity. They also witnessed vortex pairing that occurred either due to the jet-column mode or shear layer mode instability. For the jet-column mode of instability, the most stable vortex pairing took place at *St = 0.85* along with the generation of a sub-harmonic frequency. Ho and Huang[4] showed an interesting phenomenon of 'collective interaction' of the vortices in their experiment, in which they strongly forced the mixing layer at a sub-harmonic of the most amplified frequency. Moreover, they categorized the frequency into four modes based on the number of vortices involved in pairing. In mode I, when the frequency was near the most-amplified peak frequency, the vortex pairing was not observed for a long distance. When the frequency was sub-harmonic of the peak frequency in mode II regime, the response frequency (or the vortex passage frequency) near the jet exit was doubled as compared to the forcing frequency in conjunction with the vortex pairing at a fixed position. Similarly, in mode III and IV, response frequency became three and four times the forcing frequency respectively, and the same number of vortices were found to participate in vortex pairing. Ho[5] also conducted a linear stability analysis and proposed a 'subharmonic evolution' model based on the global feedback mechanism. This model postulates this said phenomenon along with the local instability process. According to the global feedback mechanism the vortex pairing at the downstream is responsible for acoustics perturbations, which would travel upstream to affect the evolution of coherent structures. These upstream propagating acoustic perturbations cause the generation of a fundamental frequency which in turn becomes responsible for the vortex pairing at the downstream. Ho and Nosseir[6] also confirmed this acoustic feedback mechanism in their study. Later on, Laufer and Monkewitz[7] analyzed the forced jet experimental data of Kibens[8] and found that the model was also valid for free jets as well. Other earlier works [9-11] reported the details about sub-harmonics and its effect as available in literature.

The study on vortex structures and their circulation, of the impulsively started incompressible jets, was carried out by Gharib *et al.*[12] and Mohseni *et al.*[13] using piston/cylinder arrangement in a water tank. They also observed a leading vortex ring with a trailing jet that was formed for higher stroke ratios while a single vortex ring formed for smaller stroke ratios. This transition was found to occur near the stroke ratio 4. This study is related to earlier discussed works in a sense that the vortex passage frequency gets doubled as compared to the forcing frequency applied at the inlet. Gharib *et al.*[12] designated the second coherent structure formed near the jet exit as trailing jet. However, this was significantly different as compared to the study of continuously forced jets at the inlet. In a forced jet, shear layer continuously feeds the vortex rings with the vorticity, unlike the piston-cylinder configuration that produces impulse jets. Here, the 'forcing' is the controlled perturbation applied at the jet exit so as to control the vortex structure formation and their interactions. The forcing can be applied using loudspeaker in the experiments or including some time-dependent function for velocity fluctuations in the numerical simulations. Recently, Gohil *et al.*[14] also witnessed the generation and pairing of the leading vortex and the trailing jets in their simulation of the low Reynolds number forced jet perturbed at *St*= 0.2. However, to the best of the author's knowledge, there is no study available in the literature that investigates the formation and the pairing/merging of *trailing vortex* with the *leading vortex* in a forced jet with coflow.

Since the present work aims to quantify the trailing and the leading vortices based on the amount of energy and circulation carried in them, we invoke the tools like modal decomposition and vortex tracking method for data analysis. Modal decomposition method like Proper Orthogonal Decomposition (POD) is used to decompose the fluctuating flow field and its vortical structures into different spatial orthogonal modes based on the fluctuating turbulent energy while Dynamic Mode Decomposition (DMD) is helpful in the decomposition of the flow field based on the frequency of the vortical structures. Similarly, vortex tracking method has been used to track the vortex characteristics like circulation, vortex generation, till the domain exit. We have further extended the study of parametric variations on vortex dynamics to have better insight about the control of jet evolution in different situation. The parameters like – turbulence intensity, momentum thickness, coflow temperature, coflow intensity and jet Reynolds number are varied. Also, the investigation includes the effect of hot coflow on the vortical structure in the cold forced jet, as this particular phenomenon is the backbone of Moderate or Intense Low-oxygen Dilution (MILD) combustion [15-19] in which relatively cold fuel jet burns in the hot diluted oxidizer to improve the combustion efficiency. Hence, term wise analysis of vorticity transport equation is carried out for the hot coflow case to explain the dynamics which is otherwise difficult to explain through other methods.

The organization of the paper is as follows – Section 2 specify the details of numerical methodology and the computational domain. The validation of the numerical code is carried out in Section 3, whereas Section 4 presents the results of simulations carried out on the forced jet with a coflow on a set of varying parameters. Section 5 discusses the summary of the results followed by the conclusions in Section 6.

## 2. Numerical Methodology

*2.1. Governing equations and solver details*

The Pencil code[20], an open source code is invoked and modified as per our needs to solve the fully compressible conservation equations of mass, momentum, and energy, which are recast as:

$$\frac{\partial \rho}{\partial t} + \nabla \bullet \rho \vec{u} = 0 \tag{1}$$

$$\rho \frac{D\vec{u}}{Dt} = -\nabla p + \nu \left( \nabla^2 \vec{u} + \frac{1}{3} \nabla \nabla \bullet \vec{u} + 2S_{ij} \bullet \nabla \ln \rho \right) \tag{2}$$

$$\frac{\partial T}{\partial t} = -\vec{u} \bullet \nabla T + \frac{1}{\rho c_v}(H - C + \nabla \bullet (K\nabla T) + 2\rho \nu S_{ij} \otimes S_{ij}) - (\gamma - 1)\nabla \bullet \vec{u} \tag{3}$$

and the equation of state is $p = \rho \frac{\Re}{W} T$.

The continuity equation solves for the density in which $\rho$ denotes density, $t$ is the time and $\vec{u}$ represents the velocity vector. The momentum conservation equation solves for the velocities in all the three directions. Here, $p$ is the static pressure and $\nu$ is the kinematic viscosity. From energy conservation equation, $T$ (temperature) is solved where $H$ is heating (equal to zero in the current simulations), $C$ is the cooling (equal to zero in the present simulations), $K$ is the thermal conductivity, $\gamma$ is the specific heat ratio of the fluid and the rate-of-shear tensor looks like:

$$S_{ij} = \frac{1}{2}\left( \frac{\partial u_i}{\partial x_j} + \frac{\partial u_j}{\partial x_i} - \frac{2}{3}\delta_{ij} \nabla \bullet \vec{u} \right) \tag{4}$$

Where $S_{ij}$ is traceless and $\delta_{ij}$ is Kronecker delta. Right-hand side terms in the energy equation include heat convected, heat removed or added to the system, heat diffused (heat conduction), work done on fluid due to fluid shearing and compressibility effects. The energy equation has been used here for simulating the case of the jet with hot-coflow where the temperature and density change occurs across the domain.

The code is written using finite difference scheme. The Runga-Kutta third-order scheme is used for explicit temporal calculations while the spatial discretization invokes sixth order accurate central difference. Third order RK-scheme reduces the amplitude and phase error and to some extent allows longer time steps. The sixth order central difference scheme has very less dissipation and is computationally very economical compared to spectral methods. Thus these equations can produce accurate solutions [20].

*2.2. Methods for Vortex Structure Analysis*

For the identification and study of the vortex structures and its dynamics, various techniques are employed to extract the vortical structures from the flow field.

*2.2.1. Proper Orthogonal Decomposition (POD)*

The POD is a mathematical tool that helps in identification of the dominant flow structures. Kosambi[21] carried out the early work, followed by the work on turbulent flows by Lumley [22]. The methodology used in the current work takes several snapshots of the instantaneous flow field as proposed by Sirovich [23]. It further calculates and sorts the POD modes based on the respective fluctuation kinetic energy (spatial) which is proportional to its eigenvalue. The fluctuation kinetic energy forms an auto-covariance matrix, calculated from the fluctuating component of the instantaneous velocity. One can refer to the published literature[24-28] for the formulation and further details related to the POD. Further, for variable density (in turn temperature) flows, both fluctuating velocity and temperature fields are taken into consideration, while the solution of POD energy matrix (a combination of kinetic energy and thermal energy [29-31]) provides the eigenvalue of the system.

*2.2.2. Dynamic Mode Decomposition (DMD)*

The important phase information in POD gets lost as it is based on the second-order statistics. Hence we have used DMD to overcome this issue. This method also takes several snapshots as input but provides with the dominant flow structures that still contain the temporal information of the same. The generated modes, also called dominant modes, are based on the dominant frequency (temporal) in the whole domain. The problem studied in this work is non-linear. Hence this method (according to the work of Peter J. Schmid [32]) produces structures of a linear tangent approximation to the underlying flow. The formulation of DMD can be further referred to work by Peter J. Schmid [32-34].

*2.2.3. Vortex Tracking Method for Circulation*

To gain in-depth knowledge of the vortices and their interactions in different situations, we track the variation of its circulation throughout the traversed path of the vortices while they undergo phenomena like vortex pairing, etc. The analysis of Gharib *et al*. [12] for the impulsively starting jets is extended here for the case of forced jets with vortices being continuously fed with vorticity. The circulation is calculated based on the iso-vorticity contours of the vortex under study. The value of the minimum contour level of vorticity is chosen such that it distinguishes the vortex being tracked from the free shear layer of the jet (Gharib *et al*. [12]) and also from other nearby vortices. Hence the minimum contour level would tend to vary for different simulation conditions, and later Section 4 provides further discussion on this. Since the unperturbed jet exit velocity along the centerline ($U_{c,0}$) remains constant with time, the vortex circulation ($\Gamma$) has been non-dimensionalized as $\Gamma/(U_{c,0}D_0)$. The non-dimensional circulation and area ($A/D_0^2$) are shown against non-dimensional time and distance, where the non-dimensional time $\tau_n$ describes as:

$$\tau_n = \frac{t - t_0}{T_p}, \tag{5}$$

Here, $T_p$ is the time period between two consecutive vortices. The parameters like minimum contour level are chosen at an approximate value $\tau_n = 0.8$ to maintain the consistency across various cases for this analysis. As the centerline velocity gives a fair idea about the location of the vortices, the variation of jet centerline velocity with respect to time easily allows computing $t_0$ and $T_p$. However, the past literature [3, 14] for studying vortices in the jet has used the streamwise velocity signals. The velocity at its peak value triggers the vortex generation and hence, the time instant at which this peak occurs is denoted by $t_0$ (see Fig. 1). In some cases, we study the circulation against the distance (*x/D*) which is non-dimensionalized by the diameter (*D*) of the jet.

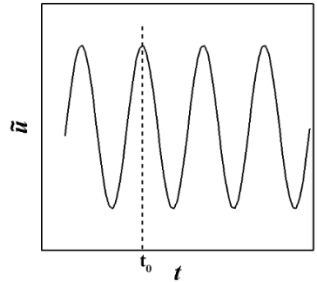

FIG. 1. Variation of velocity signal concerning the time at the jet inlet.

*2.2.4. Q-criterion and Vorticity Transport Analysis*

One of the common and an efficient method [14, 35-37] to visualize vortical structures is Q-criterion. Essentially, positive Q-criterion means the region where rotational (or vorticity) magnitude dominates over strain-rate magnitude [38]. We invoke this approach in the present analysis. On the other hand, vorticity transport equation helps to understand the behavior of the vortex as it traverses downstream. Term wise analysis of this equation for the flow field sheds light on vortex dynamics involved in the given case [39,40]. The equation describes as:

$$\frac{D\vec{\omega}}{Dt} = (\vec{\omega} \bullet \vec{\nabla})\vec{u} - \vec{\omega}(\vec{\nabla} \bullet \vec{u}) - \frac{\vec{\nabla}p \times \vec{\nabla}\rho}{\rho^2} + \upsilon \nabla^2 \vec{\omega} \tag{6}$$

The terms on the right-hand side are vortex stretching (1st), gas expansion (2nd), baroclinic production (3rd) and viscous dissipation (4th), respectively. The second and the third term play a crucial role in the case of jet-in-hot-coflow, as there are temperature and density variations in the flow field.

*2.3. Computational domain and boundary conditions*

The simulations utilize the rectangular domain with the Cartesian grid. The chosen domain measures - 16D x 12D x 12D, as depicted in Fig. 2. The grid has been refined to the scale of Kolmogorov length to account for the subtle turbulent structures that may form at some downstream distance from the jet exit. For this, the Kolmogorov length scale is determined using Boersma *et al.* [41]'s method (1998) as:

$$\eta/D = [1/(0.15 \text{Re}^3 B_u^3)]^{1/4} * x/D \tag{7}$$

where $\eta$ is Kolmogorov length scale and $B_u$ is the centerline velocity decay rate of 5.75 (the value is in accordance with the result obtained in the later Section 3 – Validation). From this we get $\eta/D = 0.0024 * x/D$. The resolution achieved in the present computations correspond to:

$$\begin{aligned} \delta x &= 1.3\eta \quad at \ x = 16D \\ \delta x &= 2.0\eta \quad at \ x = 10D \\ \delta x &= 3.0\eta \quad at \ x = 6.5D \end{aligned} \tag{8}$$

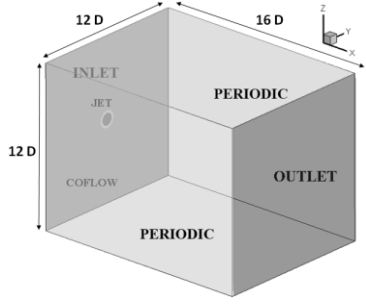

FIG. 2. Schematic of the computational domain with boundary conditions

One can note that the grid resolution is of the order of Kolmogorov scale. Hence it is capable of capturing the interaction of coherent structures as well as the transition to turbulence of the jet towards the end of the domain. The grid is fine enough to resolve small-scale structures, and that minimizes the error due to grid resolution as well.

Boundary conditions play a very critical role in case of direct numerical simulations where we resolve at the smallest dissipative scales. Since the code is higher order accurate, boundary conditions should also be able to provide equally high-order precision results. Therefore, non-reflective NSCBC (Navier-Stokes Characteristic boundary condition) boundary condition becomes an obvious choice, and we use the same in our simulations. This type of boundary condition is suitable for a transient and compressible code where the code solves the Navier Stokes equation at the boundaries [42].

### 2.3.1. Inlet Boundary Condition

At the inlet, we invoke the NSCBC boundary condition with the formulation introduced by Yoo *et al.* [43], Lodato *et al.*[44]. Therefore, every time-step solves the characteristic equations for different primitive variables [44]. The following sub-sections describe the inlet velocity and temperature profiles used in the present study.

### 2.3.1.1. Inlet Velocity Profile

The hyperbolic tangent profile is used to provide mean inlet velocity of the jet

$$U_0(r) = 0.5\left(1 + \tanh\left[\frac{R}{4\theta}\left(\frac{R}{r} - \frac{r}{R}\right)\right]\right) \qquad (9)$$

Where $U_0$ is the mean axial jet velocity, $R$ is the jet radius and $r$ is the distance between any point of choice at inlet face from the jet axis and $\theta$ is the momentum thickness of the free shear layer of the jet. The advantage of using this profile is that it is valid for high $R/\theta$ values and can also be applied to the region close to end of the potential core of the jet. Several past researchers [45-51] have used this. This condition is extended for the jet in a coflow [50] as:

$$U_0(r) = \frac{1}{2}(U_j + U_\infty) - \frac{1}{2}(U_j - U_\infty)\tanh\left[\frac{R}{4\theta}\left(\frac{r}{R} - \frac{R}{r}\right)\right] \tag{10}$$

where $U_\infty$ is the velocity of the coflow and $U_j$ represents the velocity of the jet. Additional terms have been added to the above profile to apply the 'jet column mode' perturbation at the inlet. Since the disturbance is added only to jet, equation (10) is rewritten in the form:

$$U(r,t) = U_{0,j}(r) + U_\infty + U_{forced,j}(r,t) + U_{turb,j}(r,t)$$
$$U_{0,j}(r) = \frac{1}{2}(U_j - U_\infty)\left(1 + \tanh\left[\frac{R}{4\theta}\left(\frac{R}{r} - \frac{r}{R}\right)\right]\right) \tag{11}$$
$$U_{forced,j}(r,t) = U_{0,j}(r)\sum_m A_m \sin\left(\frac{2\pi St_m U_j t}{D} - m\theta\right)\left(\frac{2r}{D}\right)^{|m|}$$

where $U_{0,j}(r) + U_\infty$ is the velocity profile of jet in a coflow, $U_{forced,j}$ is the forced perturbation added to the system, $U_{turb,j}$ is the turbulent disturbance added at the jet inlet. Here $St_m$ denotes the *Strouhal number*, $D$ is the diameter of the jet, $A_m$ is the amplitude of the forced perturbation which is fixed at 15 percent of the mean inlet velocity throughout this study, $m$ is the mode of perturbation which is zero in the varicose case, $\theta$ is the azimuthal direction of $\vec{r}$.

*2.3.1.2. Turbulence Generation*

To replicate the turbulent inlet condition in a practical system, the methods related to generation and introduction of synthetic turbulence are quite popular in numerical simulations. In this study, the turbulence is produced separately in an isothermal box with periodic boundary conditions, in which a forcing function is used to provide white noise at the inlet. When the box turbulence simulation becomes isotropic and statistically stationary, the extracted data from the isothermal box containing relevant information is fed to jet inlet in the main simulation, as reported in literature [52]. The present approach differs from the usual methods where artificial disturbances are introduced directly into the main simulation without having it to become statistically stationary. In the current study, the turbulence with an intensity of 2% and 5% is introduced for the case of turbulent, forced jet-in-coflow (Section 4.2). It is less than the forcing amplitude 15%, which is kept constant in this study. It is to be noted, firstly, that the forcing here refers to the controlled perturbation with a specific frequency and mode, whereas the turbulence introduced at the domain inlet, is uncontrolled with the mix of various frequencies. Secondly, forcing is applied to all the cases in this study (except validation) with the aim of getting a desired vortical structure, which in turn controls the jet evolution. Both the forcing and the turbulence results in an unstable flow.

*2.3.1.3. Inlet Temperature Profile*

In case of the jet in cold coflow, the temperature remains same throughout the inlet. In case of the jet in hot coflow, the temperature is high in coflow and lower in the jet. However, the temperature profile should be defined such that it takes into account the variable homogeneous fluid density and temperature of the compressible flow. The system is at constant pressure with the assumption of *Prandtl number* equal to 1, while the Busemann-Crocco law is used [53] to couple the temperature profile with the velocity profile. The relation used in the previous work [54] is limited to the hot jet in cold coflow. Hence, we have taken care of the required modification for the current case of the cold jet in hot coflow, and the resultant equation looks like:

$$T = T_\infty + (T - T_\infty)\left(\frac{u - u_\infty}{u_1 - u_\infty}\right) + \frac{(u - u_\infty)(u_1 - u)}{2c_p} \qquad (12)$$

Putting $u_\infty = 0$ gives back the relation mostly used for hot jet without a coflow [54-56]. For further details, the work done by Hermann Schlichting [53] on Crocco-Busemann equation can be referred, but one can note that this relation is based on the assumption that $\Pr = 1$, $R/\theta \gg 1$.

*2.3.2. Outlet Boundary Condition*

Similar to inlet boundary condition, the NSCBC boundary conditions are used at the outlet plane as suggested in the past literature [42,44].

## 3. Validation

Initially, we have validated the conventional free, round jet against the experimental results of P O' Neil *et al.* [57] in Fig. 3 for *Re* = 1030, which corresponds to turbulent jet with $R/\theta$ of 6.5, same as that in the experiment. The considered domain size is 24D x 12D x 12D (to be consistent experiments) with a grid size of 700 x 240 x 240. Fig. 3(a,b) depicts the half-width in the near-field and the far-field region of the jet. In the near-field region of the jet exit, the results are in excellent agreement with the experiment, while we observe some minor differences at further downstream (in turbulent regime). Fig. 3(g) represents the centerline mean velocity variation and the obtained velocity decay is very much similar to the past results of Wygnanski and Fiedler [58], Panchapakesan and Lumley [59], Hussein et al. [60]. The velocity decay rate constant ($B_u$) and spread rate (*S*) can be calculated from eq. 13 and eq. 14,

$$\frac{U_{c,0} - U_{cf,0}}{\Delta U_c} = \frac{1}{B_u}\left(\frac{x}{D} - \frac{x_0}{D}\right) \tag{13}$$

$$S = \frac{\delta_{1/2}}{(x - x_0)} \tag{14}$$

The virtual origin lies near *x/D* = 5. The mean value of decay rate constant ($B_u$) is 5.75 in region *x/D* = 15 to 24 while the spread rate is 0.12. The value of decay rate constant is in good agreement with the predicted values available in the literature (as given in table 1). This value of velocity decay is used in the calculation of the Kolmogorov length scale ($\eta$), subsequently deciding the grid size for simulations (discussed in an earlier section). The spread rate is also found to agree with the experimental result of 0.10 by Panchapakesan and Lumley's [59] and 0.11 by Gohil et al. [14].

| Authors | Decay Rate Constant | Experimental/Computational |
|---|---|---|
| Wygnanski et al. [58] | 5.4 | Experimental |
| Hussein et al. [60] | 5.8 | Experimental |
| Panchapakesan [59] | 6.06 | Experimental |
| P O'Neill et al. [57] | 5.6 | Experimental |
| Boersma et al. [41] | 5.9 | Computational |
| Wang et al. [35] | 5.5 | Computational |
| Gohil et al. [14] | 5.2 | Computational |
| **Present study** | **5.75** | **Computational** |

Table 1. Comparison of Decay Rate Constant with literature.

The mean velocity profiles are shown for the near field (at $x/D$ = 3, 4) as well as for far field from the jet exit at different downstream distances (at $x/D$ = 12, 14.1) in Fig. 3(c). As observed, both the results are consistent with the experimental data by P O' Neill [57], while the profile at far downstream shows that the fully developed jet profile because both the experimental and the computational plots are of the Gaussian shape.

This shape indicates the self-similarity profiles at far downstream. Further, the profiles of normal stresses (axial and radial) and shear stress are also depicted in the far turbulent region of the jet at the same downstream, as shown in Fig. 3(d-f), which is also in good agreement with the experimental results.

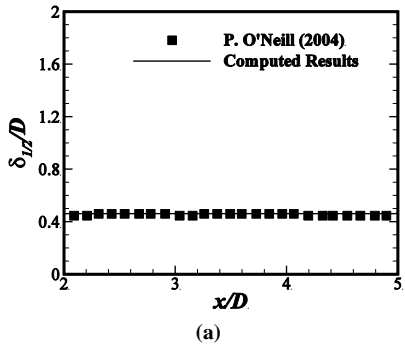
(a)

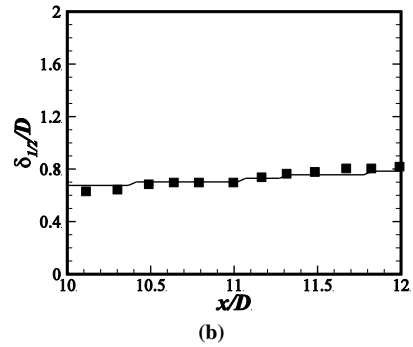
(b)

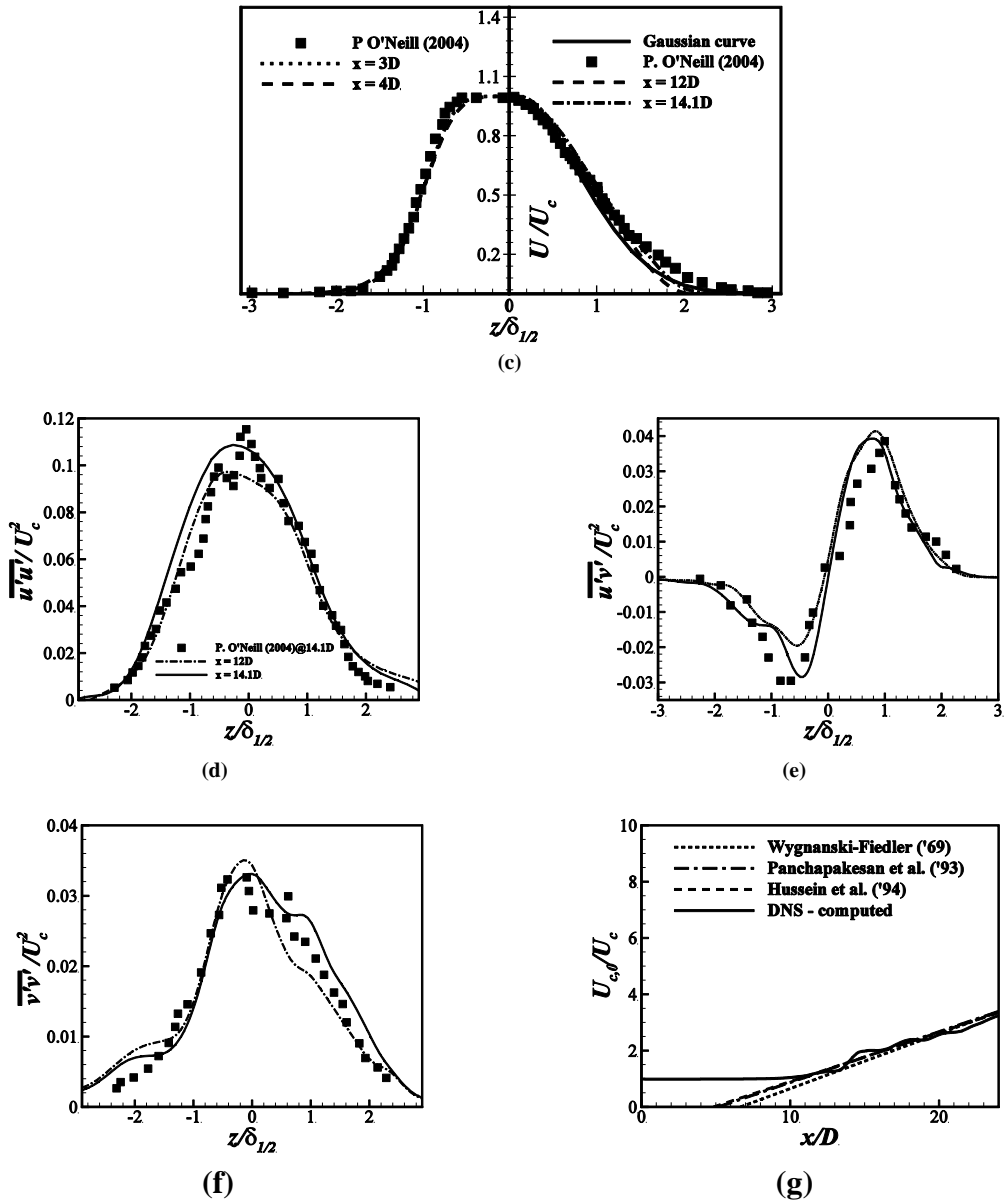

FIG. 3. Jet half - width at (a) near jet inlet region and, (b) far downstream jet region. (c) Mean velocity profiles of the jet at x = 3D, 4D and, x = 12D, 14.1D. (d), (e), (f) Reynolds stresses in the far downstream region. (g) Centreline mean velocity variation along the jet axis.

## 4. Results

*4.1. Variation of inlet forcing frequency*

In the previous studies [2-3],[14], a variation of jet-inlet perturbation frequency is found to have a significant impact on the jet evolution and the formation of vortex structures. At the preferred mode of frequency, the jet shows highest spreading and no vortical interaction, whereas at frequencies higher than the preferred mode leads to the pairing of the vortices. When the applied frequency of the inlet perturbation is less than the preferred mode, trailing edge vortices tend to form. These changes in vortex size and its dynamics need to be understood for the jet in coflow as well.

In this case, the jet has been excited at different frequencies applied at the inlet. The varicose mode of excitation is applied at the jet inlet for a fixed $Re = 1000$. One must note that the excitation is only applied in the jet and not in the coflow.

Various frequencies are used in the study with Strouhal number ($St$) = 0.2, 0.3, 0.4, 0.5 and 0.6. Q-criterion is used to visualize the coherent structures in the instantaneous flow field for these cases. In Fig. 4, three-dimensional plots of Q-criterion colored (flooded) with vorticity are shown for different $St$. The side view in Fig. 4 shows the vortex structures along with the instantaneous velocity contours. The change in the frequency of the passage of vortices or the distance between two consecutive vortices can indicate the pairing of vortices in a given condition. Based on the distances between the consecutive vortices in Fig. 4(a), (d) and (e), jets with $St = 0.2$, 0.5 and 0.6 show pairing of vortex rings whereas jets with $St = 0.3$, 0.4 (see Fig. 4(b), (c)) do not exhibit any pairing mechanism. The jet with $St = 0.2$, exhibits an early pairing between $x/D = 6$ to 7, while in the case of $St = 0.5$ the pairing is observed at around 15D to 16D. However, on the other hand, the jet with $St = 0.6$ shows pairing near ~11D downstream distance. As far as forcing frequency is concerned, the lower value of frequency means the delayed pairing due to the large distance between the vortices, which expectedly occur in $St = 0.2$ cases. However, Q-criterion iso-surfaces show that a smaller vortex ring gets generated behind the (main) leading vortex, but the same is not true in $St = 0.6$ case. The pairing process differentiates the case of the jet in coflow with $St = 0.2$ from the cases with $St = 0.5$ and 0.6. This observation is further confirmed in FFT results as shown in Fig. 6 and Fig. 7 (discussed in the next section).

The axial mean velocity profiles at different downstream distances are also plotted to show the relative spreading of jets in various cases (Fig. 5(a), (b) and (c)). Since the jets have reached steady state, mean velocity profiles are enough to enlighten about the spreading. Fig. 5 (b) reveals that, at $x/D = 16$, the largest spread of jet is for the case with $St = 0.4$, followed by the jet with $St = 0.3$, 0.5, 0.6 and then 0.2, respectively. Further simulations of jet with $St = 0.3$ and 0.4 with extended domain (size

– 32D x 12D x 12D) depict that the spread of jet with $St = 0.3$ is more than that of the jet with $St = 0.4$ (at 30 diameters downstream distance, see Fig. 5 (c)). This is again in agreement with the results of Crow & Champagne [3] and Gohil *et al.* [14]. The above discussion confirms that an early pairing decreases the jet spread.

Also, the instantaneous plots of jets corresponding to $St = 0.3, 0.4$ exhibits an early transition to turbulence (sub-structures like ribs are visible) when compared to other cases. Hence it can be said that the jet spreading is due to the early transition to turbulence. From the perspective of mixing of the jet with its surrounding, as reported by Hussain [61], mixing means the acquisition of three-dimensional random vorticity. Since turbulence involves generation of random vorticity, mixing should happen better in a perturbed jet with $St = 0.3$ and $0.4$. In other words, one can say that mixing happens more in these jets because of their increased spreading. Hernan and Jimenez [62] also reported that most of the entrainment takes place during the normal life of eddies (nearly 80 percent [61]) without having any pairing. The results obtained here also indicate that the pairing stabilizes the jet but reduces the mixing rather than causing an early transition to turbulence for the given conditions.

*4.1.1. Analysis of Vortical Structure Interaction in jet-in-coflow with Strouhal number 0.2 and 0.6 - FFT results*

Control of jet needs an understanding of dynamics and evolution of coherent structures, particularly how these structures form and interact with each other. From the above simulations, the intriguing fact is that as the frequency of the jet inlet perturbation decreases, the vortex pairing ceases to happen as observed at $St = 0.4$ and $St = 0.3$. However, the jet with $St = 0.2$ exhibits pairing as reported by Gohil *et al.* [14]. For the detailed study of the vortex dynamics and further to quantify this peculiarity in this particular case of the jet with coflow at $St = 0.2$, we have utilized various tools in the current work. The detailed analysis primarily focuses on the vortices and their interacting behavior under different conditions of inlet turbulence, momentum thickness, coflow temperature, Reynolds number and coflow velocity.

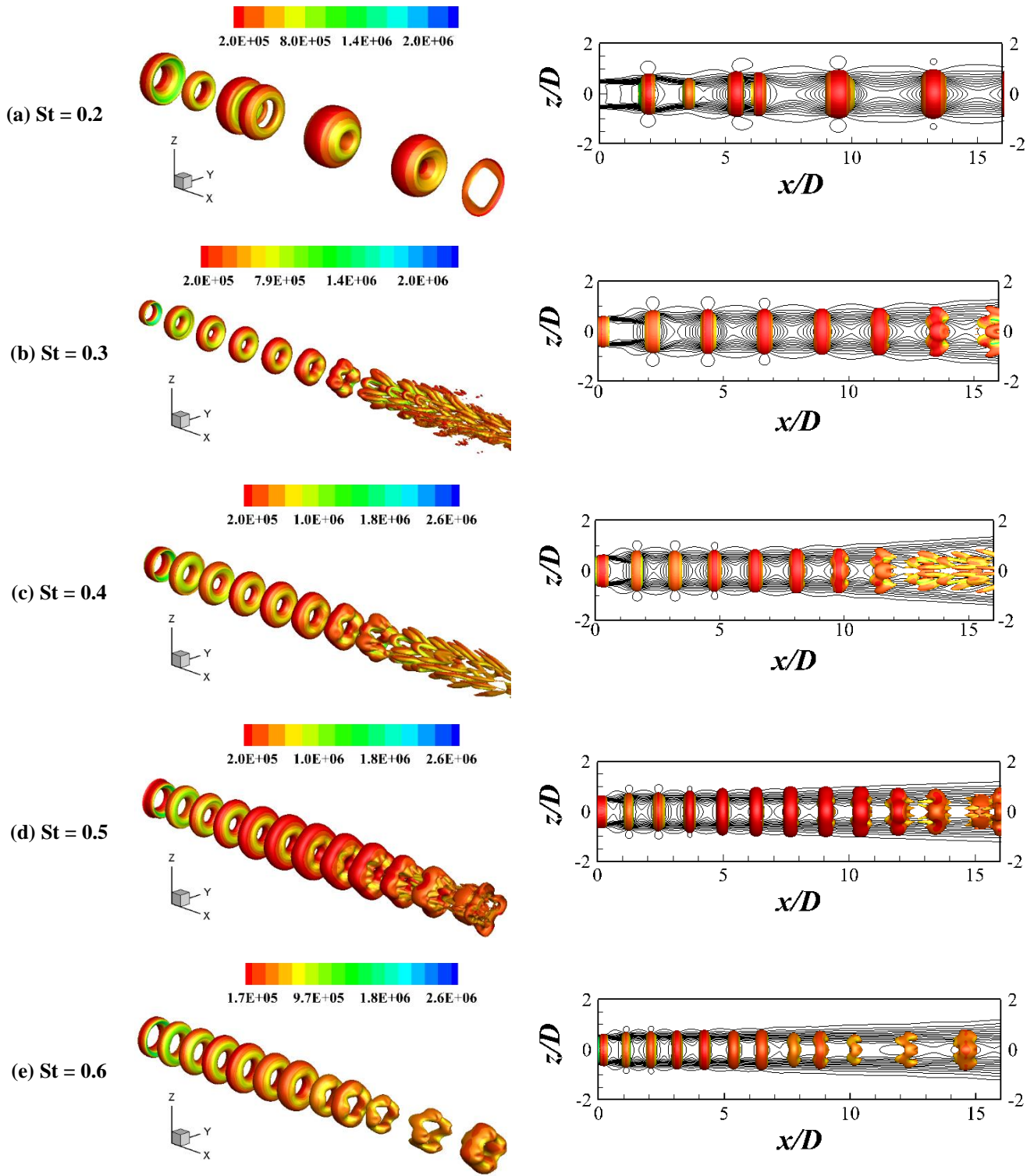

FIG. 4. (a-e) Iso-surfaces of Q - criterion flooded by vorticity magnitude for jet in cold coflow with $St$ = 0.2, 0.3, 0.4, 0.5 and, 0.6 without turbulence, in perspective and side view. Side view shows the location of vortex pairing between consecutive rings in jet with $St$ = 0.2, 0.5 and 0.6, and the jet spread in all the cases.

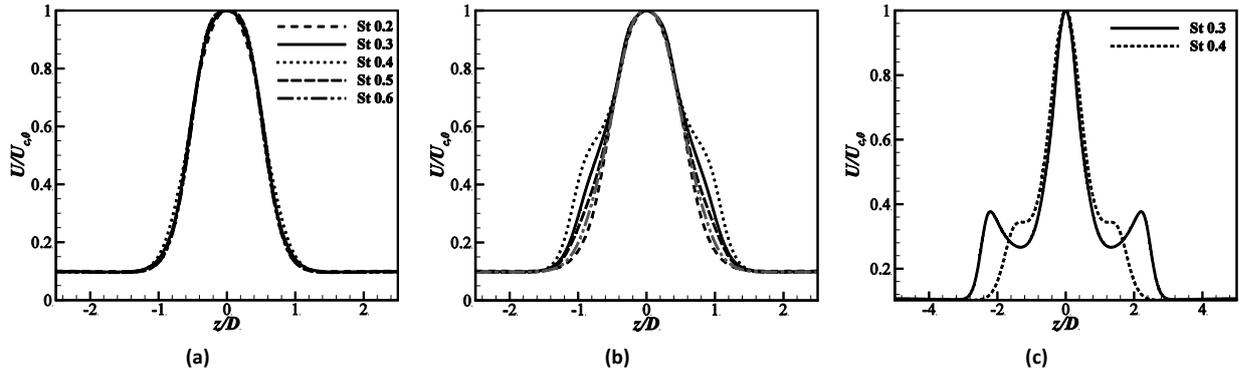

FIG. 5. Axial velocity profile of the jet at different axial locations: (a) X = 10D, (b) X = 15D and, (c) X = 30D.

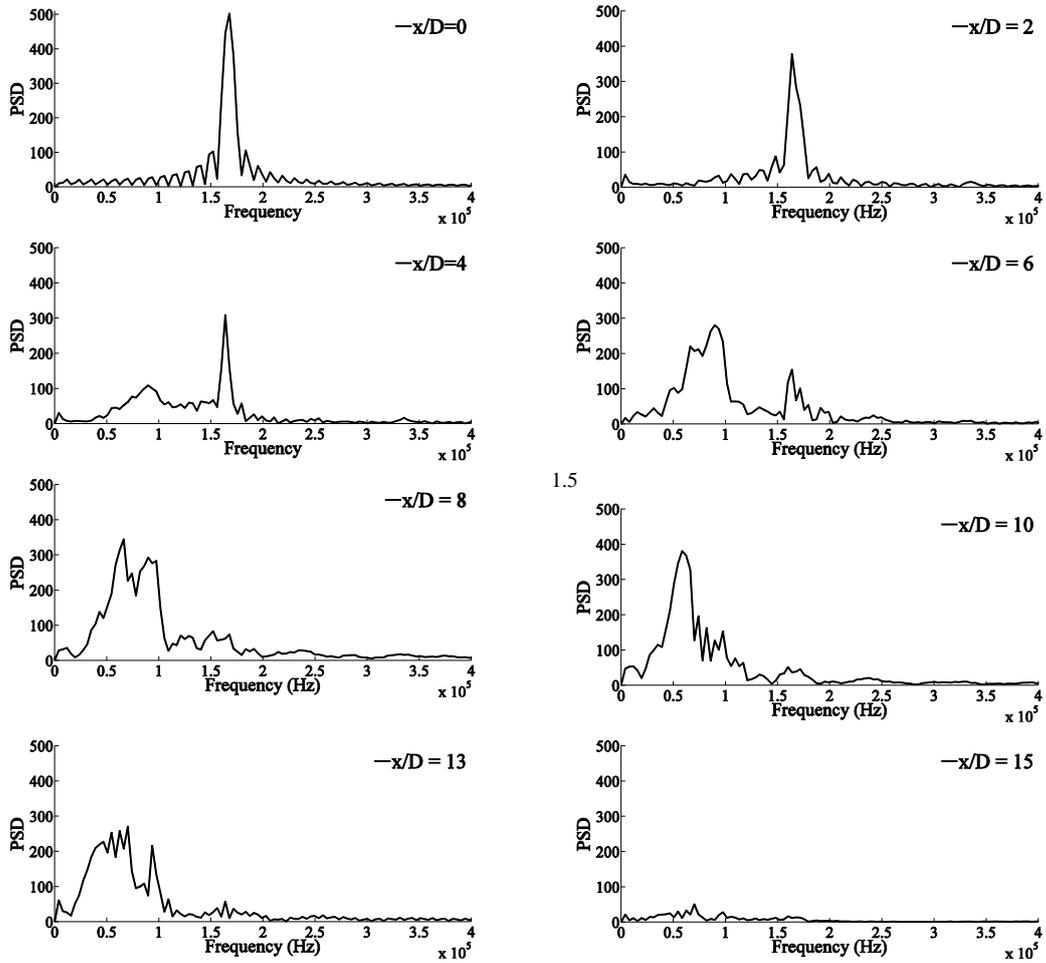

FIG. 6. FFT plots at various downstream distances for the jet in cold coflow with $St = 0.6$.

Jets with Strouhal number higher than four show stable vortex pairing as reported in the previous studies [3,14]. The distance of vortex pairing location is seen to decrease with increasing $St$ (for $St > 4$) because of the decreased separating distance between two consecutive vortices which in turn increases the local induction effect between them. Since FFT provides the basic idea of the vortex structure development and their interaction (and also used in the previous studies by past researchers [3], [14], [61-62]), it would help in obtaining the frequency corresponding to the individual vortices. Hence to show the difference in the evolution of the vortical structures of the jet in coflow with $St = 0.2$ vis-à-vis cases with higher Strouhal numbers, we perform FFT on both the jets with $St = 0.6$ & $0.2$.

*4.1.1.1. Jet with Strouhal number 0.6*

Observing the power spectral density (PSD) plots in Fig. 6, we can note the inlet frequency (fundamental frequency) in the first panel which is around $1.7 \times 10^5$ Hz. As we move downstream, this peak gets smaller and diminishes after $x/D = 8$. However, from $x/D = 4$, a sub-harmonic peak of frequency $0.65 \times 10^5$ Hz starts emerging and reaches a maximum around 10D downstream distance. Further, this peak maintains itself till $x/D = 13$, after which it deteriorates and gets mixed with the surrounding peaks which might correspond to the turbulence. This trend is significantly different from the jet with $St = 0.2$ where the fundamental peak is generated while the sub-harmonic is forced at the jet exit.

Fig. 4(e) shows the side view (span-wise direction) of the instantaneous plot. Here, vortex pairing occurs between 6D - 7D downstream distances. As we move further downstream, the paired structure appears to be present which corresponds to the sub-harmonic peak in the PSD plots. The result of this case is in agreement with the FFT analysis conducted by Zaman and Hussain [3] on the jet with a range of Strouhal number 0.7-1.0. For jets with *St* near to this range, they have observed occasional pairing.

*4.1.1.2. Jet with Strouhal number 0.2*

FFT results for the given condition are depicted in Fig. 7, where the first panel ($x/D = 0$) exhibits the input frequency at the jet inlet. It is clear as we move further downstream, the generation of some higher harmonics peaks takes place. Also, one can notice that the peak corresponding to $0.55 \times 10^5$ that exists from the jet exit, gets further strengthened as we move downstream (until $x/D = 10$). Moreover, the lately generated smaller peak corresponding to higher harmonic, like – $1.1 \times 10^5$, gets strengthened (at $x/D = 3$) and maintains itself till $x/D = 4$, subsequently, weakens rapidly after $x/D = 5 - 6$ (starts strengthening afterward). A similar trend is visible for second and other higher harmonics. The results are similar to the FFT results of a jet without coflow by Gohil *et al.* [14], Ho and Huang [4]. The nature of FFT curve for this case is different from other Strouhal number cases due to the presence of trailing jet and leading vortices as reported for free jet case [14]. Hence, this case needs to be further studied, and the authors have performed a detailed investigation of the same and explained in the following sub-section.

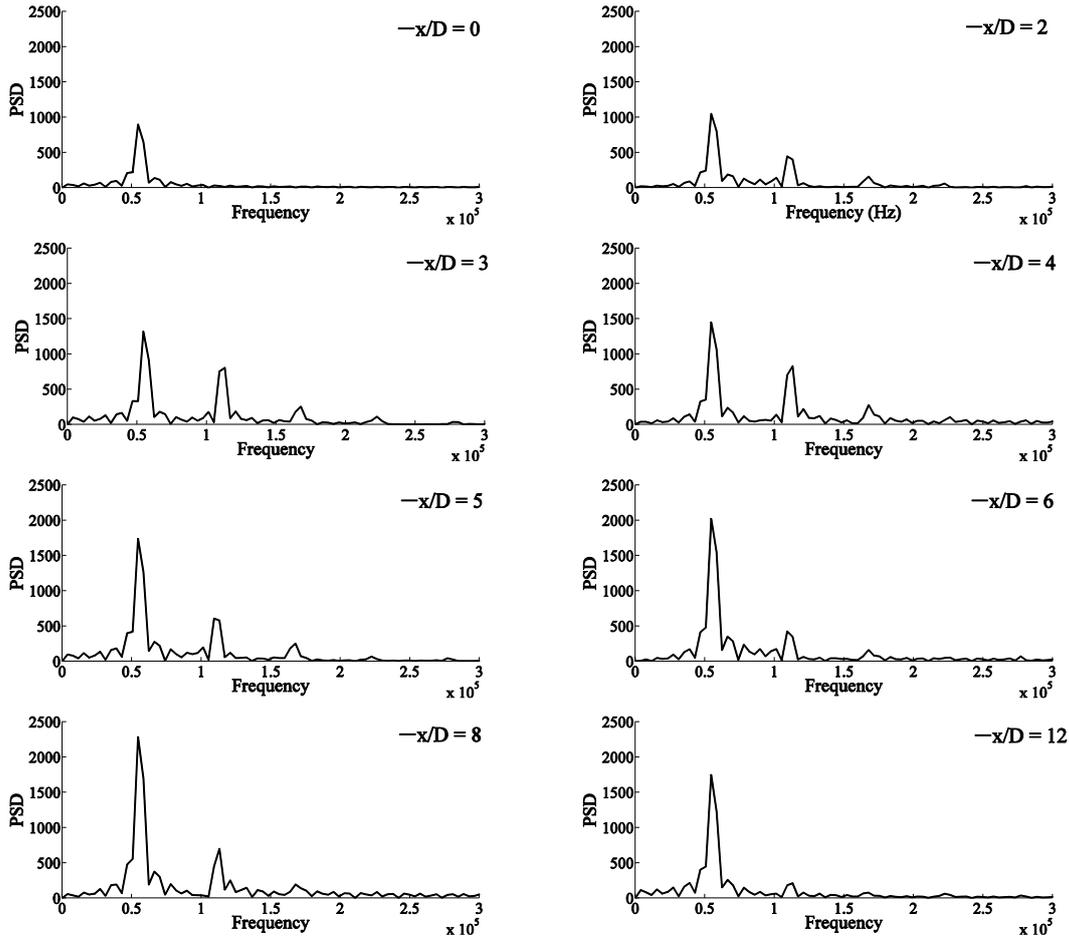

FIG. 7. FFT plots at various downstream distances for the jet in cold coflow with $St = 0.2$.

*4.1.2. Forced Jet at Strouhal number 0.2*

*4.1.2.1. Proper Orthogonal Decomposition*

The FFT results in the previous section shows the uniqueness of the case of jet in coflow with forced perturbations at $St = 0.2$. A detailed analysis needs to be carried out on the base case of $St = 0.2$, before all the other variations are imposed on it. This base case would act as a reference for all the cases with different variations of parameters. Since the modal decomposition in the spatial domain (i.e. POD) extracts the flow features on the basis of fluctuating energy, this method has been employed in the current study to get information about two kinds of vortices, namely – leading and trailing vortices, of different strengths observed at *St=0.2*. It would help in approximating the proportion of the energy between the trailing jet and the leading vortex. Hence, the highly resolved data obtained from DNS is coarsened for further analysis to reduce the computational load of 3-dimensional analysis like POD or DMD. Therefore, the grid independence test for POD has been initially carried out to confirm the size of the grid. As a part of this test, we compare the instantaneous plots of three different grids with the corresponding DNS data. Further, POD is conducted on all three of the coarsened grids to check the agreement of the POD method with the grid size used. Table 1 mentions the details related to grids chosen for the given purpose on the case-by-case basis.

Firstly, we report the Q-criterion plot flooded with vorticity magnitude at a particular instant of simulation for all the types of grids (cases - I, II, III, IV) in Fig. 8(a-d). Secondly, the velocity profiles are also compared for the same purpose (Fig. 8(e-f)). From these plots, it can be concluded based on the structures resolved, that case I, case II and case III have similar structures and shape, unlike case IV which has significantly inaccurate and thinner Q-criterion iso-surfaces and velocity profiles near the jet inlet. However, since case I and case II are computationally very expensive, case III is chosen for the detailed analysis.

| Case | Grid |
| --- | --- |
| I | 320 x 240 x 240  (DNS) |
| II | 214 x 150 x 150 |
| III | 160 x 112 x 112 |
| IV | 107 x 75 x 75 |

Table 2. Different cases for Grid Independence Test

To further quantify the grid for POD, table 3(a) shows the normalized eigenvalues against the first eight modes for all the cases. The first eight POD modes account for nearly 99 percent of total fluctuation energy. The POD results of all three cases are similar in terms of accuracy of the results, but case II and III are better than case IV regarding the resolution of the vortical structures at the inlet. Since case III can reconstruct sufficient details at low computational cost as compared to case II, we select a grid of case III (160x112x112) for further POD analysis in this work.

Having chosen the grid, we investigate the sufficiency of the number of snapshots to be considered for POD calcualtion. Table 3(b) depicts the normalized energy for first eight modes using case III grid, considering 60, 90 and 120 snapshots for the analysis. The energy of the first mode is little over-predicted and the second mode under-predicted by the case with 60 snapshots as compared to the cases with 90 and 120 snapshots. Since 90 and 120 snapshots show overlapping plots, 90 snapshots are more than enough to be considered for the POD analysis. According to the results obtained here, the detailed POD analysis is carried out considering 90 snapshots on a grid of 160x112x112.

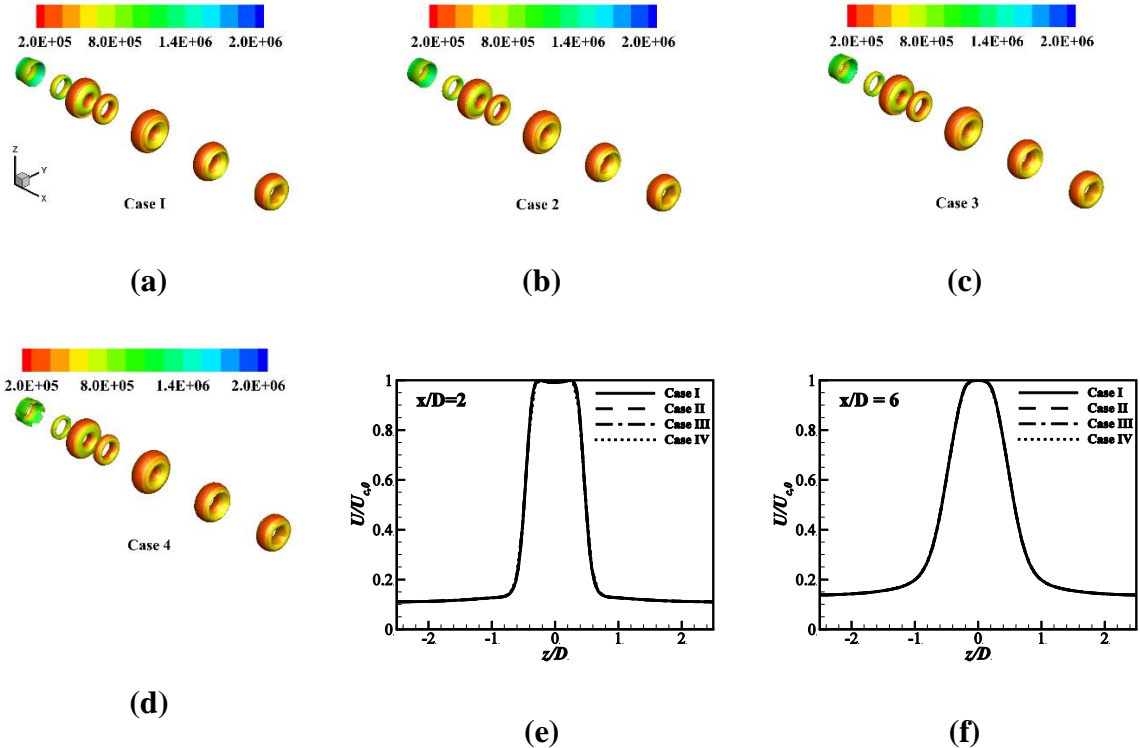

FIG. 8. Computed results on different grids (Table 2) for $St = 0.2$: (a-d) Instantaneous Q-criterion plots for case I, case II, case III and case IV respectively, (e-f) Velocity profiles at different axial distances for all the cases at (e) $x/D = 2$, (f) $x/D = 6$.

The cumulative energy plot for the given case (Case-II) is shown in Fig. 9(a), while Fig. 9(b-e) presents the Q-criterion plots of the different POD modes. Notably, the first two modes have similar energy, thereby yielding similar structures compared to mode 3 and mode 4, where the vortical structures of less energy and smaller size are more in number. However, the vortices undergoing a pairing event is captured in all the four modes, predominantly in the third and the fourth mode (Fig. 9(d-e)).

(a)

| Mode | Case II | Case III | Case IV |
|---|---|---|---|
| I | 42.99 | 43.09 | 43.01 |
| II | 42.21 | 41.92 | 42.19 |
| III | 5.95 | 5.97 | 5.96 |
| IV | 5.86 | 5.82 | 5.87 |
| V | 1.02 | 1.02 | 1.03 |
| VI | 1.01 | 1.01 | 1.01 |
| VII | 0.28 | 0.30 | 0.28 |
| VIII | 0.27 | 0.27 | 0.27 |

(b)

| Mode | Number of Snapshots | | |
|---|---|---|---|
| | 60 | 90 | 120 |
| I | 44.58 | 43.09 | 43.22 |
| II | 40.67 | 41.92 | 41.97 |
| III | 5.96 | 5.97 | 5.98 |
| IV | 5.80 | 5.82 | 5.84 |
| V | 1.03 | 1.02 | 1.03 |
| VI | 1.01 | 1.01 | 1.02 |
| VII | 0.28 | 0.30 | 0.28 |
| VIII | 0.27 | 0.27 | 0.27 |

Table 3. Results of POD grid independence for $St = 0.2$: (a) Normalized eigenvalues for case I, II and III grids, (b) Normalized eigenvalues for different number of snapshots considered on case II grid.

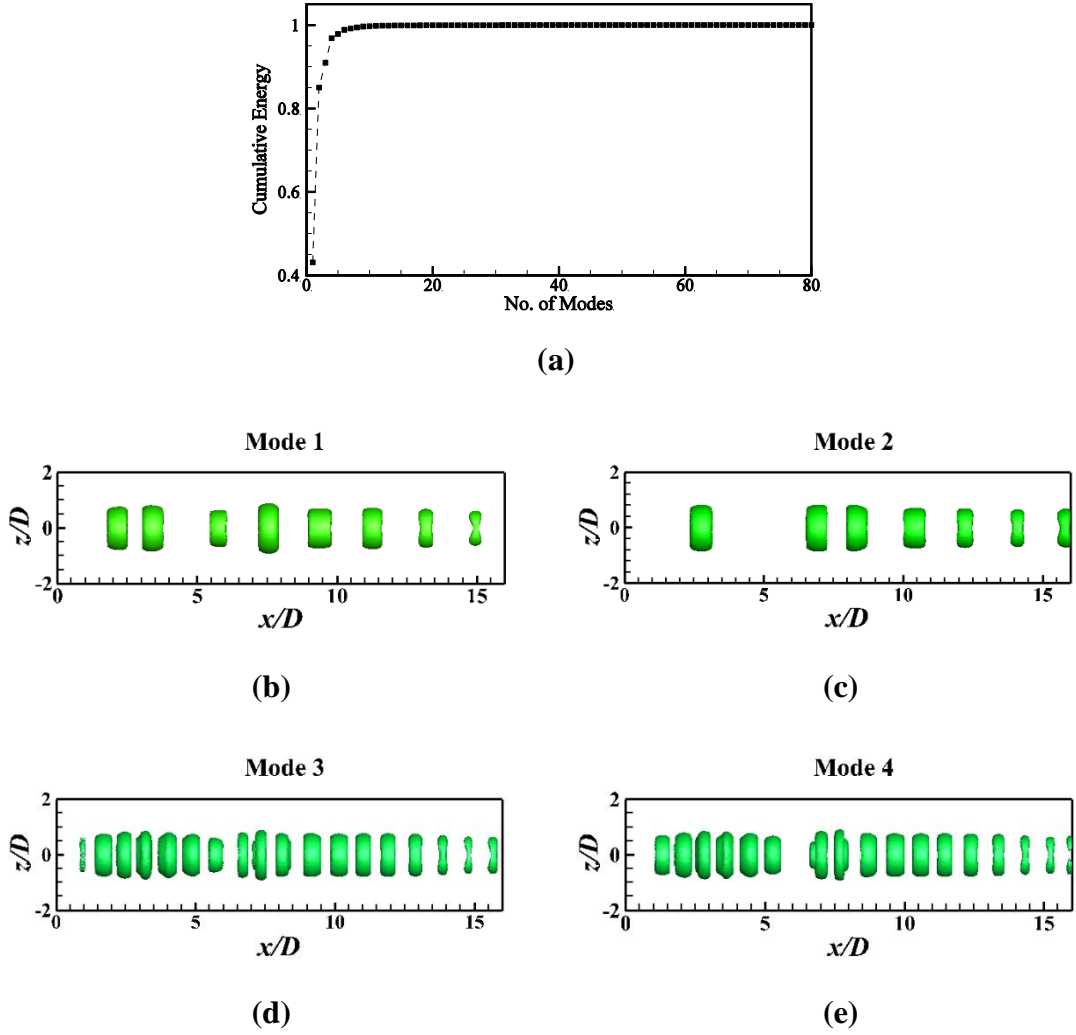

FIG. 9. POD results for $St = 0.2$ case: (a) Cumulative energy, (b-e) Iso-surface of Q-criterion plots for first four modes of the flow field.

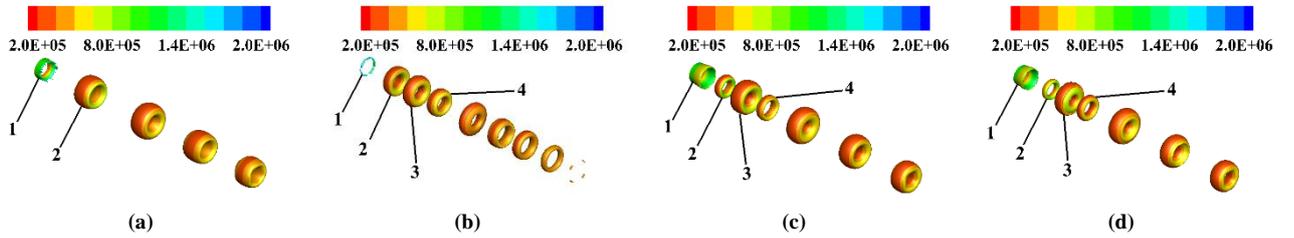

FIG. 10. Results of POD reconstruction for $St = 0.2$: (a) Modes 1 and 2 combined, (b) Modes 3 and 4 combined, (c) Modes 1-4 combined, (d) Instantaneous plot of iso-surface of Q-criterion flooded by vorticity magnitude.

The reconstruction of the flow at 90th time-step has been considered for the study of the vortical structures and their respective energy in each mode (Fig. 10). The structures corresponding to modes one and two together are presented in Fig. 10(a) and modes three and four together in Fig. 10(b).

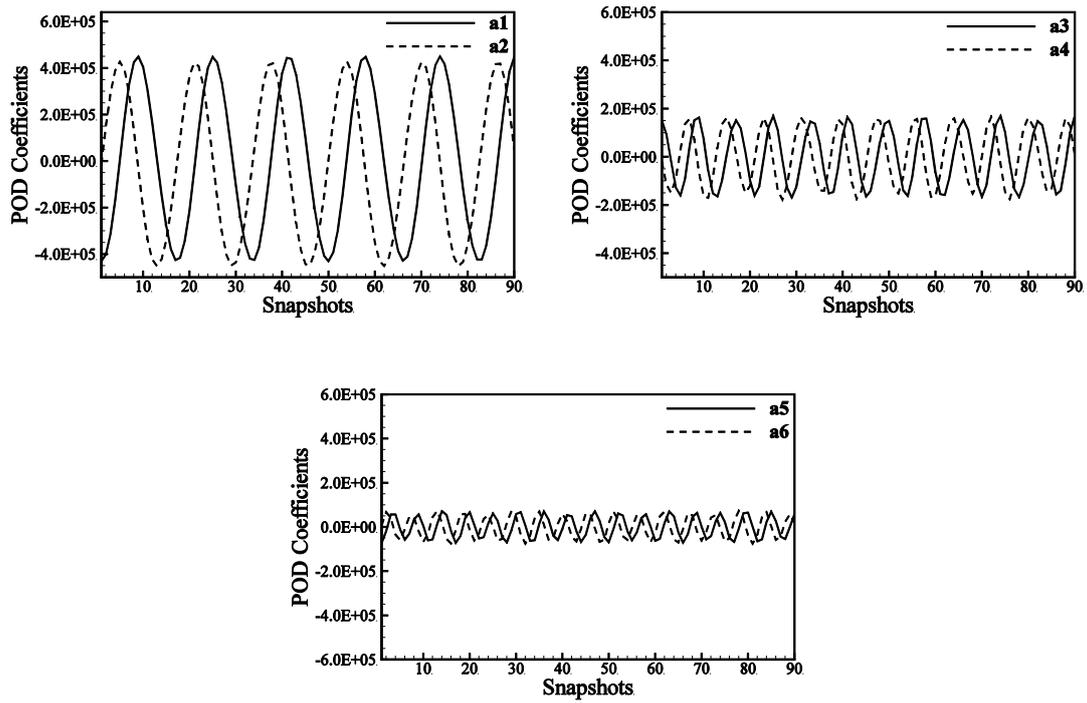

FIG. 11. Variation of POD coefficients for *St* = 0.2.

In the first one, one can observe the significant structures only. From the cumulative energy plot (Fig. 9(a)), the vortical structures shown in Fig. 10(a) possess more than 85% of the total kinetic energy (calculated from velocity fluctuations) in contrast to the structures present in Fig. 10(b). The vortical structures reconstructed from modes three and four together, contribute meager 11.8% to the total energy of (fluctuating part of) the flow. Vortical structures of Fig. 10(c) contain 96.8% of the total energy (reconstructed from modes 1 to 4) and provides a complete picture of the flow field when compared to the instantaneous data shown in Fig. 10(d). Comparing the first four vortices in Fig. 10(b-d) and the first two vortices in Fig. 10(a), it can be noticed that the second vortex in Fig. 10(a) is at the same location as of the third vortex in Fig. 10(c) and Fig. 10(d). Moreover, the third vortex in the Fig. 10(b) is also in the same position. Hence the reconstructed flow of modes one and two (i.e., Fig. 10(a)) captures a significant part of the leading (stronger) vortex; while the successive reconstructed flow using third and fourth modes (i.e., Fig. 10(b)) capture the rest of the portion of leading vortex. Whereas, out of these first four modes, the trailing vortex is captured by the third and the fourth mode only.

From this, one can conclude that the leading vortices contain more than 96.8% of the total kinetic energy of (fluctuating part of) the flow whereas trailing vortices contain nearly 11.8% of the total kinetic energy of (fluctuating part of) the flow at their respective positions at a particular time instant.

The alternate modes of the POD for this case are found to be coupled to their successive modes. As shown in Fig. 11, the variation of POD coefficients a1 and a2, a3 and a4, a5 and a6 is similar, and the odd modes are found to be lagging behind the consecutive mode by a phase difference $-\pi/2$. These pairs of modes also possess the similar amount of energy which is observed in the normalized energy plot (Table 3).

*4.1.2.2 Dynamic Mode Decomposition*

The DMD is carried out to know the coherent structures in the whole 3-dimensional space having respective dominant frequencies. The results obtained here would be reference to the jet in coflow case with turbulence. For this case, Fig. 12(a) shows the plot of $L_2$-norm of complex eigenfunctions concerning the frequency. Here, the dominant modes along with their corresponding frequencies can be noted. The first mode here corresponds to zero frequency, and second mode has frequency near to $0.55 \times 10^5$ Hz, while third and fourth modes have $1.1 \times 10^5$ and $1.65 \times 10^5$ Hz, respectively.

It is to be noted from Fig. 12(f) that the corresponding first, second and third modes are oscillatory (neither grow nor damp) as they lie on the unit circle [32], [34]. The unit circle represents the magnitude of imaginary and real values of the eigenvalues obtained from a matrix that maps between the subsequent snapshots. Modes lying within the unit circle are damped while modes lying outside the circle are amplified. Since the first mode of DMD corresponds to the mean flow field which has zero frequency, Fig. 12(a) only shows the higher, dominant modes. Fig. 12(b-c) depicts the second dynamic mode with frequency same as that of the fundamental one. In comparison with the Q-criterion plot of the instantaneous field, it is clear that the dominant coherent structures are the larger leading vortices before vortex pairing and the paired structure after vortex pairing. Here, DMD helps in identification of the coherent structures corresponding to a particular frequency (fundamental frequency in this case of mode 2) irrespective of the energy possessed by them [32]. The whole region with these dominant structures lies between $x/D = 2$ to $x/D = 13$ (after which the structures are not as dominant as upstream). This is the same region across the longitudinal length where the sub-harmonic frequency has the maximum power (near to and more than 1500) as obtained in the FFT results (from $x/D = 4.5$ to more than 12 in Fig. 7).

In mode 3 (Fig. 12(d-e)), the dominant coherent structures corresponding to the first harmonic frequency exists near to the jet exit where the vortices separate from the shear layer (from $x/D = 2$ to 5). Here, the coherent structures are smaller as compared to that of mode 2. These facts again indicate the trailing vortices together with the leading vortices are responsible for the generation of such kind of vortical structures with a fundamental frequency. These results efficiently exhibit that the region of pairing has half the frequency compared to the area just before pairing. It indicates towards the frequency of vortices encountered in the respective regions.

In FFT diagrams (Fig.7), the sub-harmonic peak at the inlet corresponds to the inlet frequency ($x/D = 0$ in Fig. 7). This particular frequency corresponds to more prominent vortical structures (leading vortices) only. As moving downstream, a smaller peak of fundamental shows up ($x/D = 2$ in Fig. 7) that indicates towards the weaker trailing vortex generated between the two leading vortices and shown as iso-surface of the Q-criterion in the third mode of DMD plot (Fig. 12(d-e)). From $x/D = 2$ onwards, both the peaks appear throughout the downstream distance. As we travel downstream to $x/D = 4$, both the peaks in FFT diagram strengthen because of the ongoing formation of the axisymmetric vortices which continue to extract energy by stretching the streamwise vortices [61]. Further downstream after $x/D = 5$, the strength of fundamental begins to decrease in contrast to the sub-harmonic peak that increases, which strongly suggests the initiation of the pairing process.

The initiation of pairing is also visible in the DMD iso-surface plots where the dominant structures of mode 3 are not visible after $x/D = 5$, and further strengthening of the structures of mode 2 takes place (Fig. 12(c) and Fig. 12(e)). Near $x/D = 8$ distance downstream of the jet exit, shows strong FFT peak of the sub-harmonic which indicates toward vortex pairing (Fig. 7). The pairing is visible in Fig. 12(c) as a form of sizeable vortical structure present at $x/D = 8$.

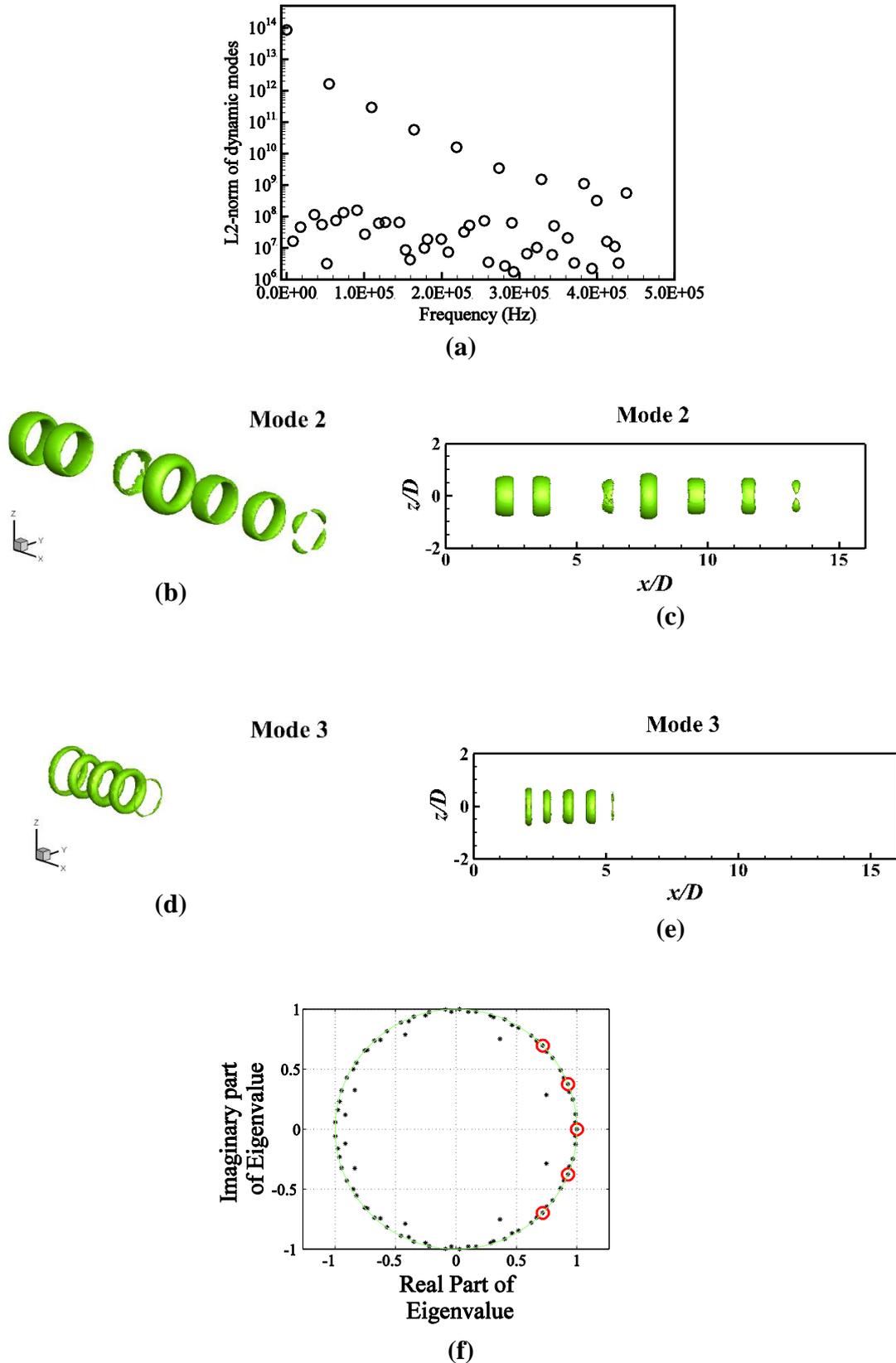

FIG. 12. DMD results for $St = 0.2$: (a) L2 norm of eigenfunction shows the frequency corresponding to the dominant modes, (b-c) Iso-surface of the second mode (fundamental) in its perspective and side view, (d-e) shows iso-surface of the coherent structures corresponding to third mode (first harmonic), (f) Unit circle diagram with first three modes highlighted.

It is clear from these results that the interaction between leading and trailing vortices, due to the pairing process, leads to the surge in the strength of the sub-harmonic peak and results in diminishing of the fundamental peak. As discussed above, the trailing vortices are the reason behind the generation of the fundamental peak as depicted by the vortex structures in the third mode of DMD. These structures are present in the region between the domain inlet and $x/D = 6$. Also, it is evident that the FFT results and DMD results complement each other, which needs further investigation as discussed in the upcoming sub-sections. Though FFT is said to provide the details regarding events occurring in the jet flow field, it cannot help in visualization of the spatial structures of a particular frequency [34].

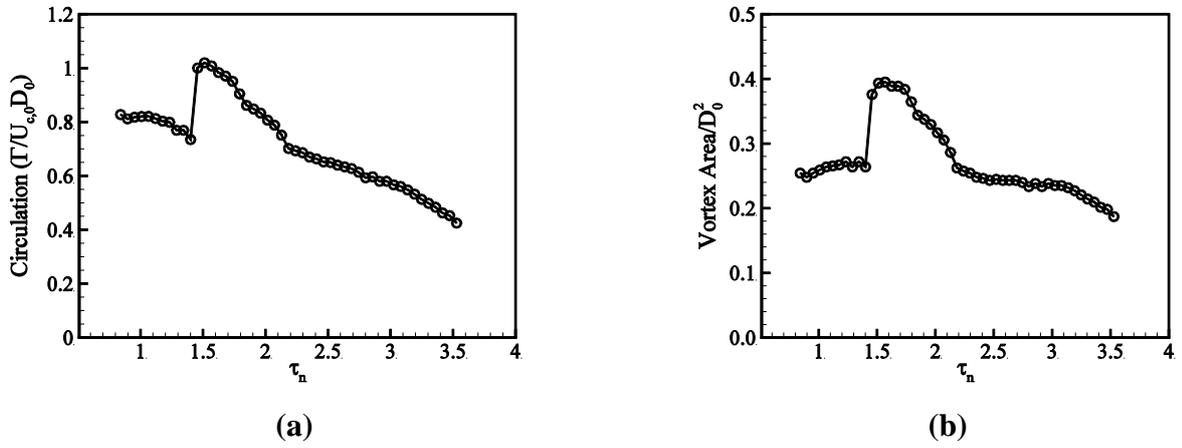

FIG. 13. Change in circulation and area of the leading vortex with time for the jet in cold coflow with $St = 0.2$.

*4.1.2.3. Circulation*

The leading vortex is tracked for the calculation of circulation (shown in Fig. 13(a)) as it traverses downstream in this case. It forms a major part of this study as it quantifies the evolution of vortical structures in the jet-in-coflow. The circulation curve shows a sudden jump at non-dimensional time $\tau_n$ = 1.4 due to the pairing event (Non-dimensionalization of time is according to equation 5).

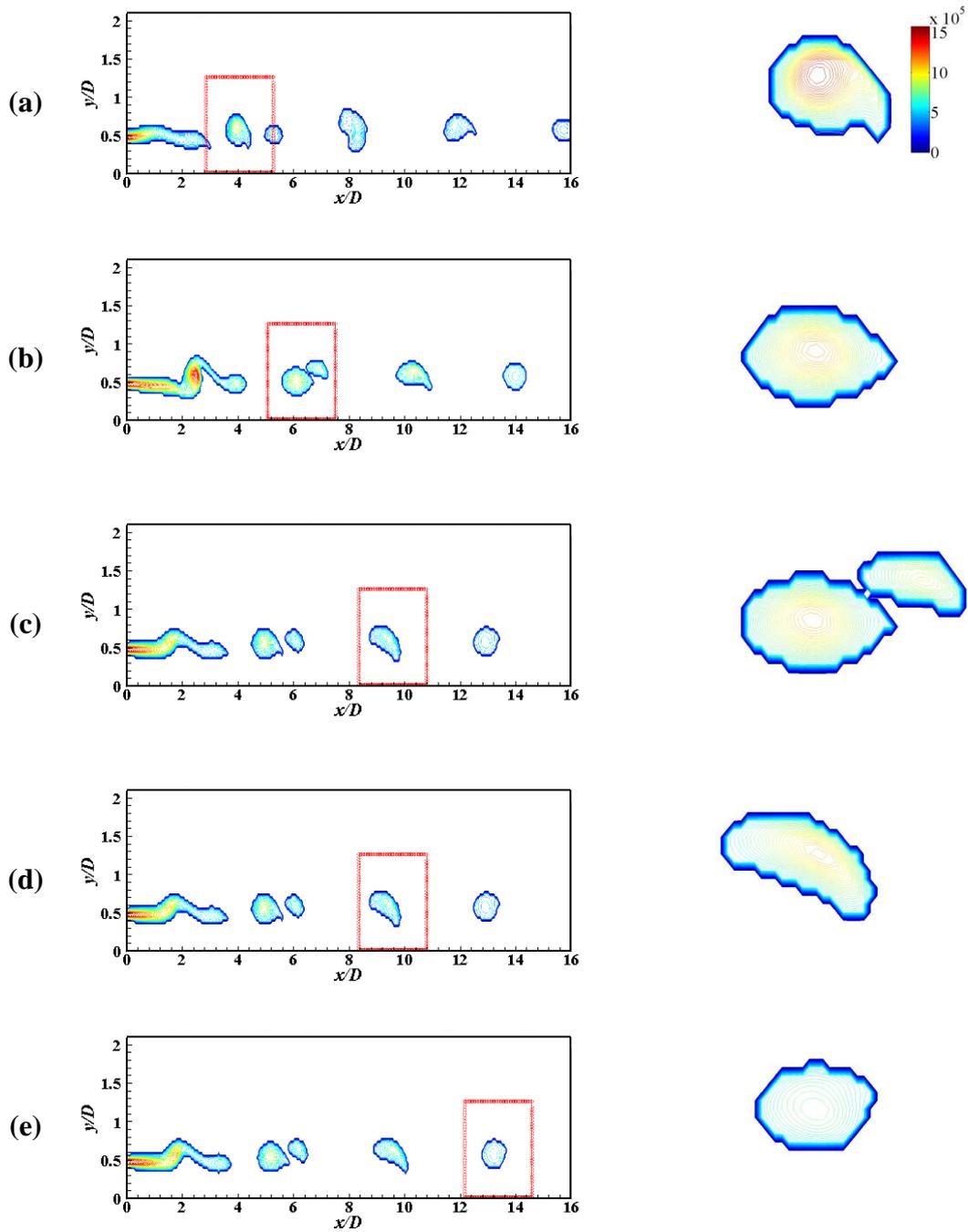

FIG. 14. Snapshots of vortex at different time instants ($\tau_n$) for $St = 0.2$: (a) 0.85, (b) 1.4, (c) 1.45, (d) 2.18, (e) 3.76.

Before pairing, there is a decrease in circulation (Fig. 13(a)) opposite to the trend observed for the vortex area (Fig. 13(b)) concerning time. The decrease in circulation can be attributed to the viscous dissipation of the vortex. For further details, the vorticity contours of the whole flow field and the vortex (enclosed by a rectangular boundary) under study are reported at significant time instants in the first column and just the highlighted vortex in the second column of Fig. 14. The lowest iso-contour value of vorticity for the vortex segregation from the jet shear layer is considered to be $5.7 \times 10^5$ rad/s. We have obtained the given value after observing the vorticity contour of the flow field at different time instants. At this minimum value of vorticity, the leading vortex stands separated from the jet shear layer from the inlet side as well as the shear layer connecting the leading vortex to the trailing vortex. In both Fig. 13 and Fig. 14, it can be concluded that the separation of the leading vortex from the shear layer at this minimum value of iso-vorticity contour level takes place somewhere around $\tau_n = 0.85$ and the pairing process starts before $\tau_n = 1.4$. It should not be confused with the fact that the vortex has separated from the shear layer. It merely indicates the connection of vortices with each other and the shear layer as well by any vorticity contour lines (until they undergo vortex pairing). During the pairing process, the attachment of the vortices can be observed just after $\tau_n = 1.45$; whereas this process of "leap-frogging" (refer to Zaman and Hussain [4] for more details) continues for some more time and finally the pairing process ends around $\tau_n = 2.18$. At the later time instants, around $\tau_n = 3.1$, the vortices dissipate at a faster rate and the circulation and the area curves sharply decrease after that.

Another observation from Fig. 13(a) is 35% increase in the non-dimensional circulation after vortex pairing. This increment appears to be 26% of the total circulation, which means that the trailing vortex carries 26% of the total circulation at the time of vortex pairing. The leap-frog motion is also visible during the pairing process; however, the paired vortical structures disappear (in the jet free shear layer) due to the viscous dissipation at the downstream locations. Hence, we have considered this case as a reference case for the other cases.

Becker & Massaro [1] showed that the wavelength for a perturbed jet with the resonant frequencies is related to the Reynolds number by: $\lambda/D = 43/\sqrt{Re}$, whereas the vortex pairing occurred at a distance downstream of the inlet. In the current case, the approximate wavelength is found out to be $\lambda = 3.6D$. The wavelength shows the measure of the distance between the center of the two leading vortex cores generated because of the perturbation at the inlet, and their frequency is same as the exciting frequency. Since the excess vorticity of the leading vortices 12-14 causes the generation of the trailing vortices, the wavelength calculation does not account for these trailing vortices. The vortex pairing between trailing and leading vortices is found to occur at almost $1.65\lambda$ from the inlet.

*4.2. Variation of turbulence intensity (St = 0.2)*

*4.2.1. Inlet turbulence intensity – 2%*

In this section, we introduce the turbulence with the intensities of 2% and 5% along with the forced perturbations at *St* = 0.2 at the inlet of the jet with cold coflow. The turbulence has been added in a manner to resemble a realistic and practical systems (refer to section 2.2 for the introduction of turbulence at jet inlet). To get insight on the effect of turbulence on the energy-carrying structures, we carry out POD analysis on the simulation data. Table 4 depicts the normalized energy plots concerning the number of modes. Here, the energy of the first five modes of POD accounts for ~54% of the total energy of the flow field while twelve modes account for ~ 80% of the energy for both the cases. To further study the structures that dominate the jet flow, we have only considered the first five modes for comparison. Another point to be noted is that the successive modes here too have similar energy like the reference case (intensity level-0%), but they are not coupled, unlike the reference case. In the case of the jet with a turbulence intensity of 2%, two successive modes have almost similar energy whereas in the case of the jet with a turbulence intensity of 5%, first four modes have almost similar energy after which the individual energy of each mode drops. Fig. 15 represents the comparison of first three POD modes for all three cases considered herein. The other POD modes show turbulent structures and hence not shown here.

It is noteworthy to mention that only the first two modes have vortical structures while the rest exhibits only the turbulent structures in the case with a turbulence intensity of 2%. As observed in Fig. 15, the pairing process appears to be absent in the Q-criterion plot of vortical structures for the three most dominant POD modes of jet perturbed with the intensity of 2%, unlike the reference state where the third and the fourth mode capture the predominant pairing of vortices.

| Mode | Turbulence Intensity Level | | |
|---|---|---|---|
| | 0% (ref.) | 2% | 5% |
| I | 43.09 | 16.05 | 13.19 |
| II | 41.92 | 15.58 | 12.10 |
| III | 5.97 | 8.92 | 11.67 |
| IV | 5.82 | 8.21 | 10.54 |
| V | 1.02 | 4.81 | 5.84 |
| VI | 1.01 | 4.67 | 4.97 |
| VII | 0.30 | 4.03 | 4.78 |
| VIII | 0.27 | 3.49 | 3.94 |

Table 4. The normalized eigenvalues for the jet in cold coflow at $St = 0.2$ with different turbulence intensities.

Fig. 16(a) depicts different frequencies present in the flow field, and the L2-norm of complex eigenfunctions (DMD results) confirms the same. Noticeably, the mean (0Hz) and sub-harmonic ($0.55 \times 10^5$ Hz) frequencies are observed to be dominant as first and second modes, respectively (Fig. 17(a)). After that, the successor modes are third and fourth mode with turbulent frequencies of $0.14 \times 10^5$ Hz and $0.40 \times 10^5$ Hz, respectively (Fig. 17(b-c)). Smaller (trailing) vortices generated in this turbulent case are captured in mode five along with the leading vortices resulting in the fundamental frequency of $1.1 \times 10^5$ Hz (Fig. 17(d)). Notably, the region in the computational domain with sub-harmonic frequency captures the larger coherent structures – leading vortices, and also the event of vortex pairing (between leading and trailing vortices). Whereas, the fifth mode with fundamental shows smaller structures – organized more uniformly near the jet inlet and consisting of more turbulent-like streamwise vortical structures near the domain exit (Fig. 17(d)). The structures of mode 5 (fundamental) in the region of vortex pairing ($x/D = 6 – 8.5$) are not that profound as in the case of mode 2 (fundamental). Neither the two vortices (– leading and trailing) have combined to form a paired structure, nor they undergo leap-frog motion of vortex pairing. Also, it appears as if the two vortices (leading and trailing) approach sufficiently nearer to each other during pairing, they act more like one larger vortical structure, which possesses a sub-harmonic frequency. In mode 3 and 4, the turbulent frequency of $0.14 \times 10^5$ Hz and $0.4 \times 10^5$ Hz is originated from the turbulent part of the jet-in-cold coflow (Fig. 17(b-c)) near the outlet of the domain. The presence of dominant turbulent modes forms a significant difference with the reference case.

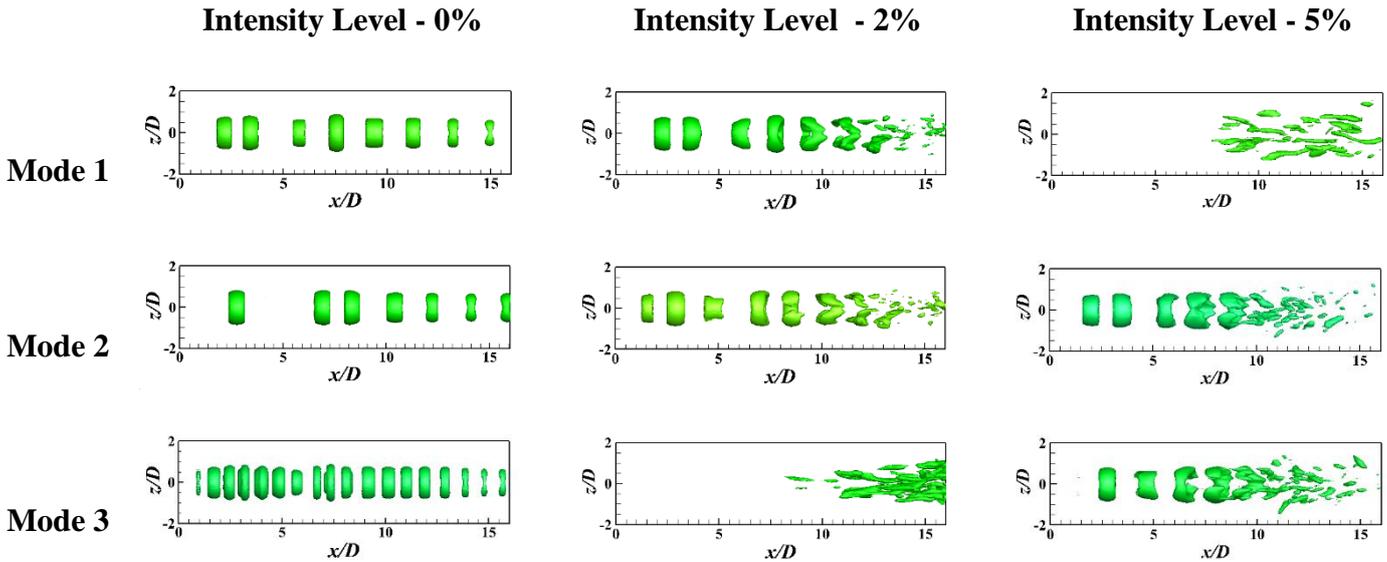

FIG. 15. Iso-surface of Q-criterion of POD modes for the jet in cold coflow at $St = 0.2$, with different turbulence intensities.

*4.2.2. Inlet turbulence intensity - 5%*

As observed, the first four modes of the current case have the (fluctuating) kinetic energy an order of magnitude higher than the rest of the modes (Table 4). This is different from the previous case where the jet with the turbulent intensity of 2% is mostly dominated by the vortical structures present in the first two POD modes. The trend for the rest of the modes shows more gradual decline afterward. Fig. 15 confirms the existence of the most dominant mode, which happens to be a completely turbulent structure and present in the second half of the domain; while the second and third modes show the coherent vortical structures. Again in this case, smaller trailing vortical structures in the pairing process are not dominant enough, and hence, they are not part of the five most dominant POD modes, though sixth mode show some portion of it in the pairing region (not shown). Further, a major difference with the reference case observed here is that the successive modes are not coupled to each other similar to the jet in coflow with inlet turbulent intensity of 2%.

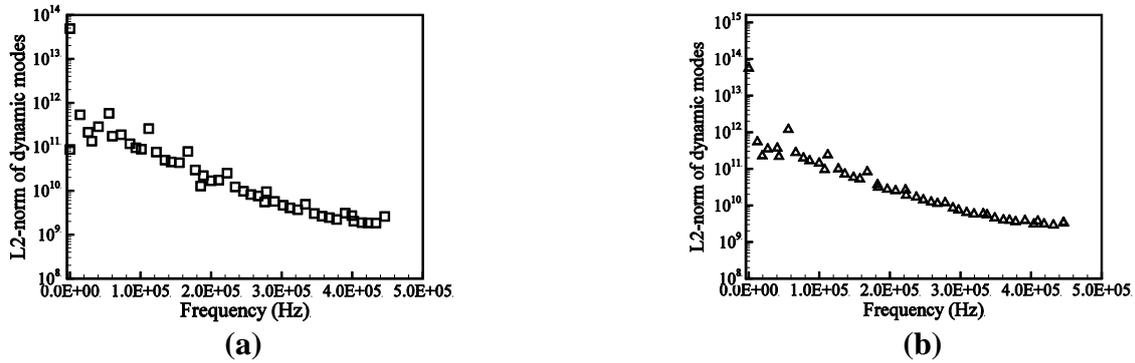

FIG. 16. L2 norm of the eigen function shows the frequency corresponding to the DMD modes for jet in coflow with turbulence intensity of (a) 2%, (b) 5%.

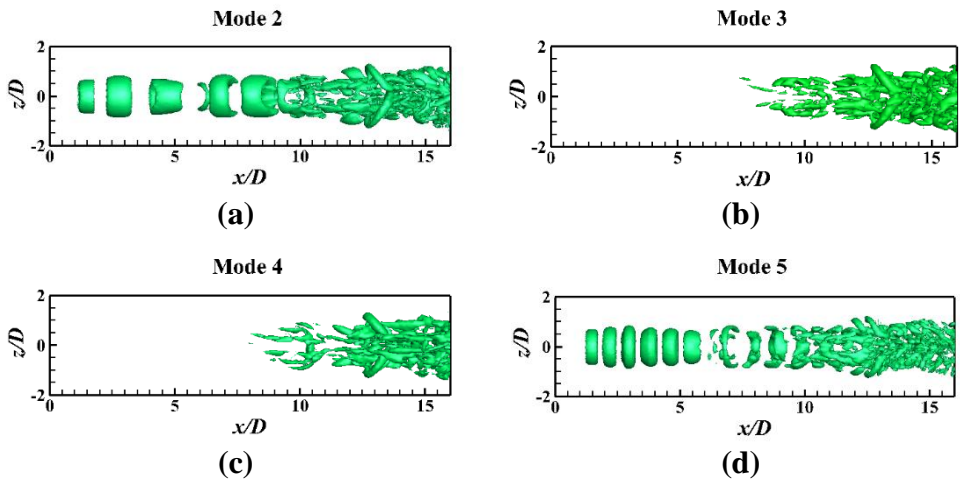

FIG. 17. Iso-surface of the four DMD modes of the jet in coflow with turbulence intensity of 2% show, (a) region corresponding to the mean frequency as 1st mode, (b) region corresponding to the frequency of pairing as 2nd mode, (c) region corresponding to the frequency due to turbulence as 3rd mode, (d) region corresponding to the frequency of region upstream of pairing(which is same as at the inlet) as 4th mode.

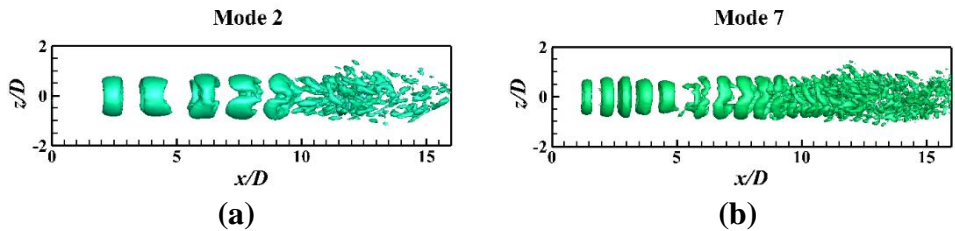

FIG. 18. Iso-surface of the four DMD modes of the jet in coflow with turbulence intensity of 5% show, (a) region corresponding to the pairing frequency as 2nd mode, (b) region corresponding to the frequency at a place upstream of pairing (which is same as at the inlet) as 7th mode.

Noticing the $L_2$-norm of the eigenfunctions of DMD results, shown in Fig. 16(b), we observe that first two dominant modes with frequencies of 0 Hz and $0.55 \times 10^5$ Hz (sub-harmonic) are same as the previous cases of turbulence intensities, i.e. 0% and 2%. The region corresponding to the sub-harmonic frequency, as depicted in Fig. 18(a), is the region of larger vortical structures or the leading vortices. According to $L_2$-norm of eigenfunctions (Fig. 16(b)), the region corresponding to fundamental ($1.1 \times 10^5$ Hz) is the 7th mode (shown in Fig. 18(b)) preceded by four modes of turbulent frequencies. Essentially, the Fig. 16(b) and Fig. 18 confirm that there are structures with turbulent frequencies that are more dominant than the fundamental one.

Again the structures in the region of vortex pairing are dominant in DMD mode 7. Since this work is mostly confined to the study of the vortices and their interaction, we do not show the turbulent structures in the context of this study. The $L_2$-norm values of the eigenfunctions related to the sub-harmonic and the fundamental frequency decrease insignificantly as the turbulence intensity go up while the number of modes, consisting of turbulent structures, also increase.

Comparing the $L_2$-norm curves of all three turbulent intensity cases, one must note that the $L_2$-norm of the modes related to sub-harmonic and fundamental remains approximately constant with increasing turbulence intensity but there is a surge in the number of modes related to turbulent structures (see Figs. 12(a), 16(a), 16(b)). It shows that the structures with turbulent frequencies become more dominant as the turbulent inlet intensity increases. Comparing both the present and the previous case, DMD has emerged to be more useful than POD in this particular case (of turbulence at jet inlet) in locating the modes corresponding to structures before and after vortex pairing because both DMD as well as vortex pairing phenomenon depend on the frequency. POD results are indicative of the fact that the small (trailing) structures are not that energetic when the turbulent structures exist in the flow field; but the dominance of the frequency of these smaller structures is comparable to the frequency of turbulent structures, as shown by the DMD results.

Analyzing the data from the perspective of the POD, it is quite evident that two continuous modes in the reference case (*Re*=1000 with *St*=0.2) are found to be coupled and have a similar amount of POD (kinetic) energy. However, the introduction of turbulence at the inlet removes this coupling between the consecutive modes of the POD which now have a different amount of energy (as seen in Fig. 15). Noticeably, the turbulence starts dominating over the vortical structures (vortex rings generated by the forced perturbations) as the inlet turbulence intensity increases from 0% to 5%. In the case of 2% intensity, the vortex rings are visible in first two, most-energetic modes. In the case of 5% intensity, the POD mode with highly turbulent structures has the highest amount of fluctuating energy, followed consecutively by two modes with vortical structures.

## 4.3. Variation of radius to momentum thickness ratio (R/Θ)

As the shear-layer momentum thickness increases, the velocity gradient decreases. This in turn, decreases the strength of the vortex structures. To investigate the effect of momentum thickness on the behavior of coherent structure formation and interaction, and also the jet evolution, radius to momentum thicknesss ratio, $R/\theta$ is varied for a constant radius. Various parameters like the vorticity contours, circulation strength of the structures, their interaction time instants and their energy have been observed this set of simulations.

### 4.3.1. R/Θ = 10

The normalized kinetic (POD based) energy content obtained from POD analysis for each case of varying $R/\theta$ and has been reported in Table 5. The first two modes of the present case ($R/\theta = 10$) has the highest energy as compared to first two modes of the reference case of $R/\theta = 20$, followed by $R/\theta = 30$. However, this order reverses in the case of energy values (normalized eigenvalue) of third and fourth mode. The energy plot of respective modes reveals (Table 5) that first and second modes together carry around 88.2% energy, while the energy of third and fourth modes together is about 13.1%. This observation is similar to the discussion about (the fluctuating part of) the energy of the vortices reported for the reference case (refer section 5.2.1). Since vortical structures of all four modes of this case are similar to the $R/\theta = 20$ case, one can note (by the discussion in section 5.2.1) that trailing vortices in carrying 13.1% energy while larger leading vortices, as shown in all the four modes, have 97.8% of the energy. The Q-criterion iso-surface of different modes confirm that mode one and mode two do not have small trailing vortices unlike modes 3 and 4. Again, this observation has similarity with the reference case of $R/\theta = 20$.

The vorticity contours (flooded with vorticity) are shown at different time instants (see Fig. 19) for the given case to illustrate the existence of vortex structures disintegrated from the main leading vortices similar to the previous study [14]. Also, the additional vortical structure appears to be formed in between the two vortices due to the tearing phenomenon, though it is not that significant because of the minimum magnitude of vorticity. First six modes of the POD do not capture this structure because of the deficient energy of the structures. Hence, the results obtained here show that the coflow around forced jet causes little transfer of vorticity (i.e. weaker vortices as compared to the vortex tearing reported by Gohil *et al.* [14]) from the already paired structure to the structure of succeeding cycle that has recently undergone pairing. Since modes 3 and 4 (and not modes 1 and 2) capture trailing vortices, in particular, we separately analyze mode 3 (and not mode 4 because modes 3 and 4 are coupled) to have a comprehensive understanding of the phenomenon that takes place with varying $R/\theta$ as shown in its perspective view in Fig. 20. Notably, a trailing vortex ring initially separates from the leading vortex of the successive perturbation cycle near *x/D* = 4-5. And, then it pairs with it at around 7D- 8D vis-à-vis reference case, where the separation of trailing ring has already taken place around *x/D* = 4 (visible as a small indent on the vortex rings near to inlet in the reference case). The phenomena of this pairing process suggest the delayed formation of the trailing vortex in this case of $R/\theta = 10$. Moreover, when the trailing vortex gets separated from the leading vortex of the successive cycle, it appears to be very thin (D1 < D2) but the vortex ring diameter (or radius from the jet centerline) is larger than the leading vortex (r1 > r2). Some of the iso-vorticity contours of the extracted vortex (with a minimum value of 600000 rad/s) at different instants, as shown in Fig. 21, supports the above statement.

| Mode | R/Θ | | |
|---|---|---|---|
| | **10** | **20 (ref.)** | **30** |
| I | 45.57 | 43.09 | 42.29 |
| II | 42.61 | 41.92 | 40.38 |
| III | 4.81 | 5.97 | 6.64 |
| IV | 4.79 | 5.82 | 6.45 |
| V | 0.80 | 1.02 | 1.30 |
| VI | 0.78 | 1.01 | 1.28 |
| VII | 0.17 | 0.30 | 0.31 |
| VIII | 0.17 | 0.27 | 0.31 |

Table 5. Normalized eigenvalues of POD modes for jet-in-coflow with varying $R/\Theta$.

*4.3.1.1. Circulation*

Fig. 22(a-b) shows the variation of circulation and the geometric area of the leading vortex against the non-dimensional time, while Fig. 22(c) depicts the same against downstream distance. The minimum value of iso-vorticity contour is used to distinguish the vortices from the jet shear layer is $6 \times 10^5$ rad/s. The circulation of a vortex can be affected because of either the change in *vortex geometrical area* or the change in the vorticity contour values within the vortex or both. The vortex geometrical area variation along with the change in vorticity contour values with time and distance would help to know which factor at a particular instant or downstream distance, out of the two factors mentioned affects vortex circulation the most. Since jet with $R/\theta = 10$ only shows a peculiar hump in the circulation curve after the vortex pairing event, Fig. 22(c-d) along with Fig. 19 help in the explanation of the same.

In Fig. 22(a) the initial drop might be caused by the leading vortex as it moves inside the trailing vortex (of the previous cycle) ahead. The observed fact can be explained by the decrease in the area of the leading vortex under study (as shown in Fig. 22(b)) as it accelerates and comes nearer to the centerline of the jet just before pairing. At non-dimensional time instant of 1.2, there is a sudden jump which is an indication of initiation of the pairing process.

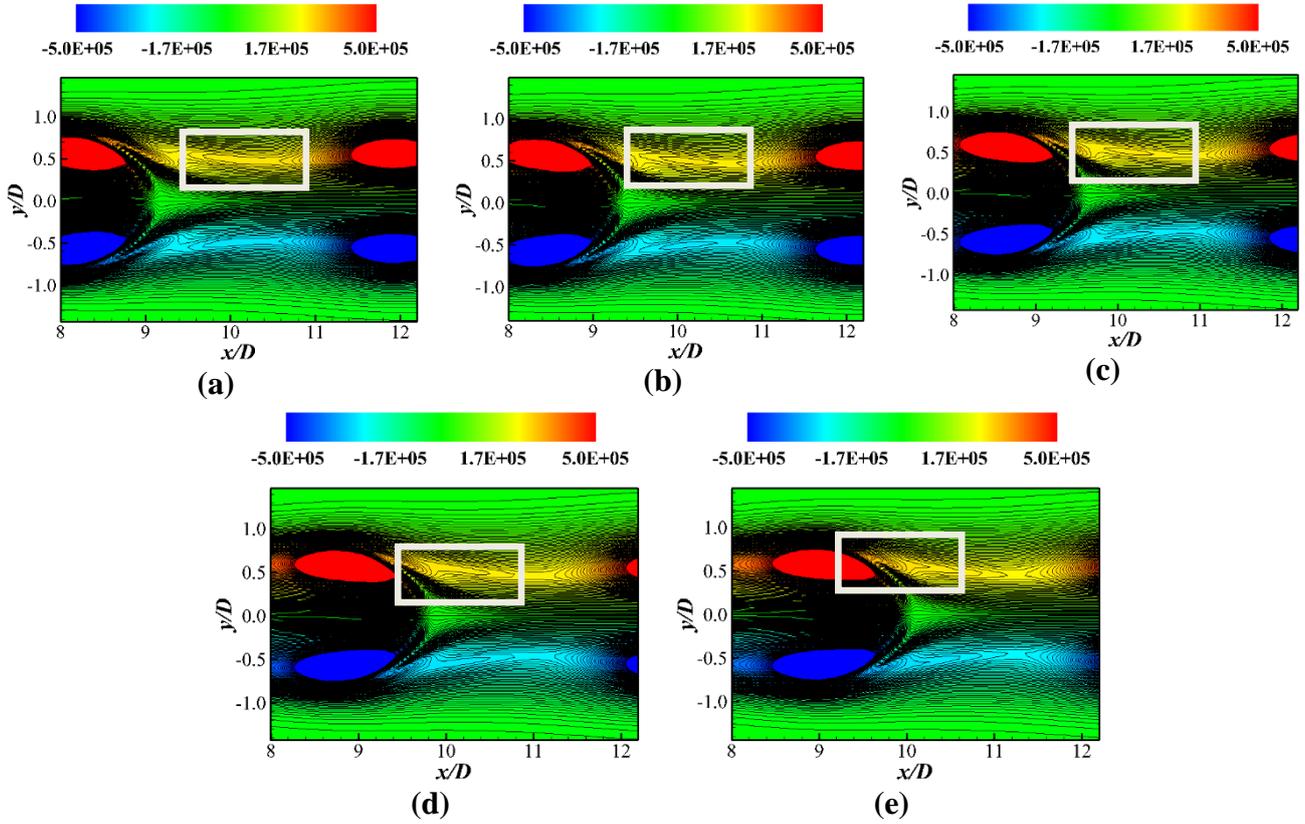

FIG. 19. Iso-contour of vorticity at consecutive time steps for $R/\theta = 10$ showing the transfer of very weak vortical structure from downstream vortex to the upstream one at $\tau_n =$ (a) 1.95 (b) 2.01, (c) 2.08, (d) 2.14, (e) 2.20.

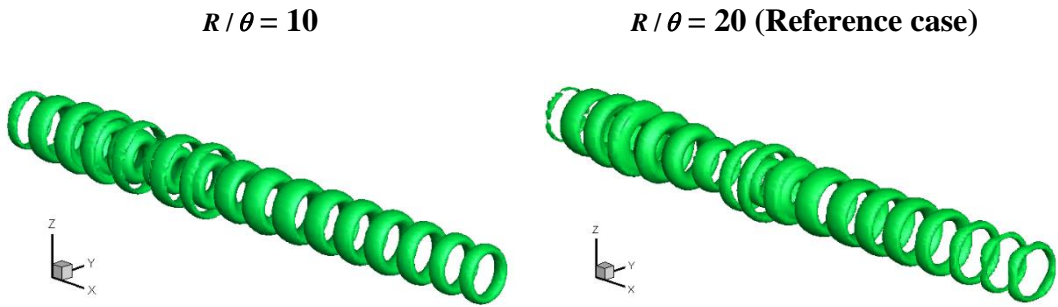

FIG. 20. Comparison of (side view and perspective view of) iso-surface of the 3$^{rd}$ POD for jet-in coflow with $R/\theta = 10$ and $R/\theta = 20$.

Near $\tau_n$ = 1.7, the pairing has ended as the graph gets straightened with a slight increase. There is an increase in circulation of leading vortex by approximately 22% due to the addition of circulation of the trailing vortex (which is an increment of 18% of the total circulation (leading + trailing)). From Fig. 22(c) and (d), the straightened patch with slight positive slope till the vertical line at $x/D$ = 9 or $\tau_n$ = 2.14 can be attributed to the pairing of torn vortex structure (as described through instantaneous contour plots in Fig. 19). This straightening instead of decay is due to this addition of some circulation from the weak torn vortex. There is no sudden increase in circulation like the pairing of leading and trailing vortices because this has been a gradual process rather than an abrupt formation of trailing vortices and an ensuing pairing process. After that, the only dominant phenomenon seems to be present is viscous dissipation. Down below the curve, near $x/D$ =13 the tearing process happens, therefore showing the decrement in circulation curve at a higher pace. This tearing is same as the one that has happened for the vortex of preceding cycle (as also seen in vorticity contour plots in Fig. 19). As observed from circulation curve, the pairing has not been as dominant as in the reference case. Further, the iso-vorticity plots of leading vortex under study confirm the above statements (Fig. 21).

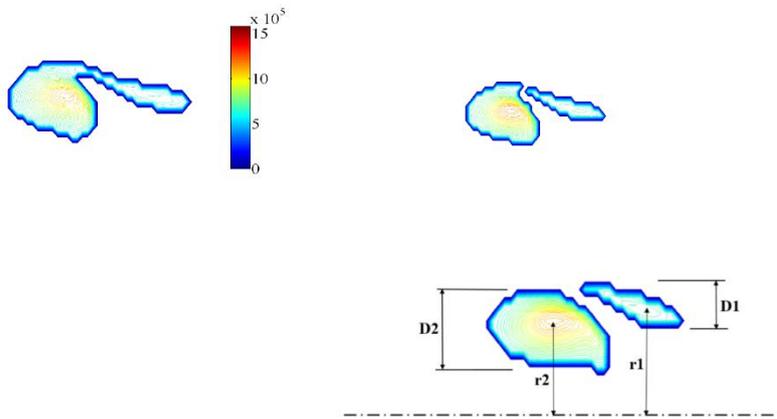

FIG. 21. Iso-vorticity contours for $R/\theta$ = 10 (showing the vortex radius - r1 and r2; the thickness - D1 and D2, of the trailing and the leading vortex rings respectively)

*4.3.2. R/Θ = 30*

In this case also, Table 5 exhibits the cumulative and the individual normalized POD energy plots against the number of modes. From the Q-criterion iso-surfaces corresponding to different POD modes, it can again be noticed that first and second modes for $R/\theta = 30$ do not show the smaller trailing vortices which is evident in the third and fourth modes, like the reference case ($R/\theta = 20$). It can also be observed from the energy of the respective modes (Table 5) that energy carried by the first and the second mode together is around 82.6% while the energy of third and fourth mode together is about 9.6%. For the same reason as in $R/\theta = 10$, it can be seen that the trailing vortices in the case of carrying 9.6% energy while bigger leading vortices have 95.7% of the energy. Also, the POD modes 1 and 2 of this case show almost identical vortical structures as that of the modes 1 and 2 of reference case.

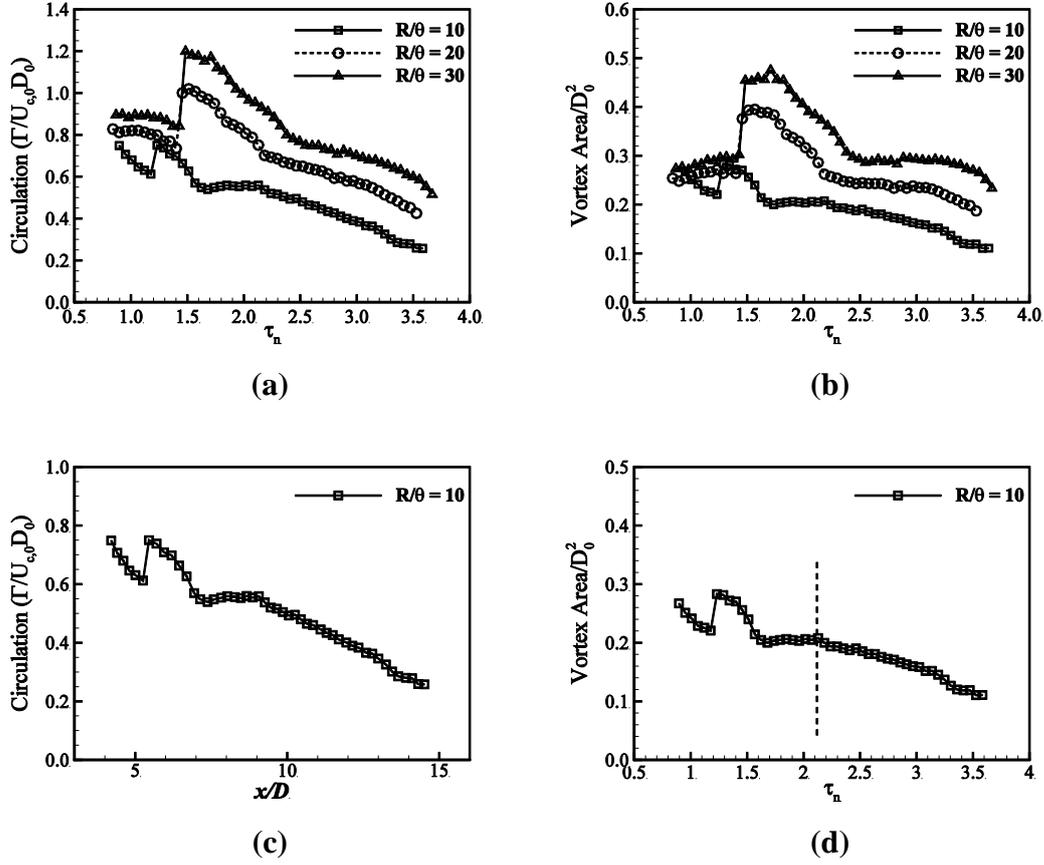

FIG. 22. (a-b) Circulation and Vortex area against time for varying momentum thickness; (c-d) variation of circulation and area of the vortex at a different downstream distance and time instants respectively when $R/\theta = 10$.

*4.3.2.1. Circulation*

The plot of circulation (Fig. 22(a)) of the leading vortex of the current case is again similar to the curve in the reference case. The minimum value of the iso-vorticity contour level used to distinguish the vortex from the shear layer is $5.3 \times 10^5$ rad/s. Similar to the case of $R/\theta = 10$ in this case also, the combined effect of a change in vortex area and variation of vorticity contour values as the (leading) vortex approaches jet centerline may attribute to the initial drop in the circulation curve before pairing. However, the sudden jump in circulation curve is higher than the case of radius-momentum thickness ratio 10, and it is due to the pairing of the leading vortex with the stronger trailing vortex, which is much stronger than both the earlier cases of thicker momentum thickness. The increment in circulation due to the addition of trailing vortices is nearly 41%, which is 29% of the total circulation (leading + trailing); even the circulation of the leading vortex itself is higher for the given case. Afterward, the circulation curve goes on decreasing with a very slight hump that shows the insignificant process of vortex tearing and second pairing - as discussed earlier. The decrease in the curve is again due to dissipation. Since the overall behavior of leading vortex is similar to the reference case, the vorticity contour plot of the vortex for this case, has been omitted.

Q-criterion plots of the reconstructed flow field for the reference case, show the energy distribution amongst the leading and trailing vortices (Fig. 10). As observed, the trailing vortices have around 11.8% of the total kinetic energy whereas larger leading vortices contain more than 97% of the total energy of (fluctuating part of) the flow (approximately eight times that of the trailing vortex) at their respective position and a particular instant. Similarly, in the cases of reduced and increased momentum thickness of the free shear layer of jet-in-coflow, the larger leading vortices are present in modes 1 and 2 of POD, while modes 3 and 4 show the smaller trailing vortical structures. From the respective POD energy curve for both the cases (see Table 5.), the energy of trailing vortices is found to increase from 9.6% to 13.1% (of the total energy of the fluctuating part of the flow) with increasing momentum thickness (from $R/\theta = 10$ to $R/\theta = 30$). This fact can also be confirmed from the circulation plots (Fig. 22(a)), where the jump in the curve due to vortex pairing happens to be maximum for $R/\theta = 10$

. The circulation of the trailing vortex while added to that of the leading vortex is maximum for $R/\theta = 30$. Hence, decreasing the momentum thickness of the shear layer increases the circulation or the energy of trailing vortex. The first few POD modes denote more dominating, energetic and large-scale structures while the higher POD modes consist of small-scale, low energy structures [24]. The first two modes (modes 1 and 2) taken together, constitute 88.2%, 85.0% and 82.6% of the total energy (of the fluctuating part of the flow) for $R/\theta = 10$, $R/\theta = 20$ and $R/\theta = 30$ cases, respectively. First few modes show the leading vortices only which means they are more dominating and energetic. Accounting the combined energy of first four modes, the energy of the leading vortex is found to have 97.8%, 96.8% and 95.7% of the total kinetic energy for the respective cases. The energy distribution strongly suggests the decrement of the proportion of the total energy of the large-scale leading vortices with the decreasing momentum thickness. The trend for the proportion of the total energy contained by the leading vortices is opposite to what is shown by the absolute values of the total energy (see Fig. 23(a)). The figure shows an increase in the energy of mode one and two; third and fourth taken together respectively, which is opposite to the trend with respect to the proportion of energy – viz., 97.8%, 96.8% and 95.7% of the total kinetic energy carried by the leading vortices for $R/\theta = 10$, $R/\theta = 20$ and $R/\theta = 30$ case.

Similarly, the rise in the circulation of the leading vortices during vortex pairing is found to increase with the decreasing momentum thickness, which is 22%, 36% and 40% for the cases - $R/\theta = 10$, $R/\theta = 20$ and $R/\theta = 30$ respectively. This jump in circulation, during vortex pairing, is 17%, 26%, and 29% when calculated concerning the total circulation of leading + trailing vortices. The proper representation of the latter values becomes possible when the vortex pairing starts. When considering the total circulation of leading + trailing vortices, the increment of 17%, 26%, and 29% interprets as the proportion of total circulation being carried by trailing vortices at the start of vortex pairing. Hence, the proportion of circulation for the leading vortices being 83%, 74%, 71% is found to decrease while that of the trailing vortices increase, with an increase in the momentum thickness ratio. Notably, the magnitude of circulation of the leading as well as trailing vortices increases on decreasing the momentum thickness (or increasing the momentum thickness ratio, see Fig. 23 (b)).

Further, Fig. 23 (c) shows the time duration of the pairing phenomenon, which shows that the vortex pairing duration increases as momentum thickness decreases (or momentum thickness ratio increases). The difference between the two-time instants - the starting and the end point - decides the pairing duration. The starting point of the pairing duration denotes the point when the leading and trailing vortices come into contact with each other as marked by a sudden jump in the circulation curve. Whereas, the endpoint is where the continuously decreasing circulation curve straightens up (slope of the circulation curve changes) as the trailing vortex ultimately merges with the leading vortex. Iso-vorticity contours shown in Fig. 21 depict these pivotal time instances of the start and the end of vortex pairing, while Fig. 22 exhibits the circulation curve for all the respective cases.

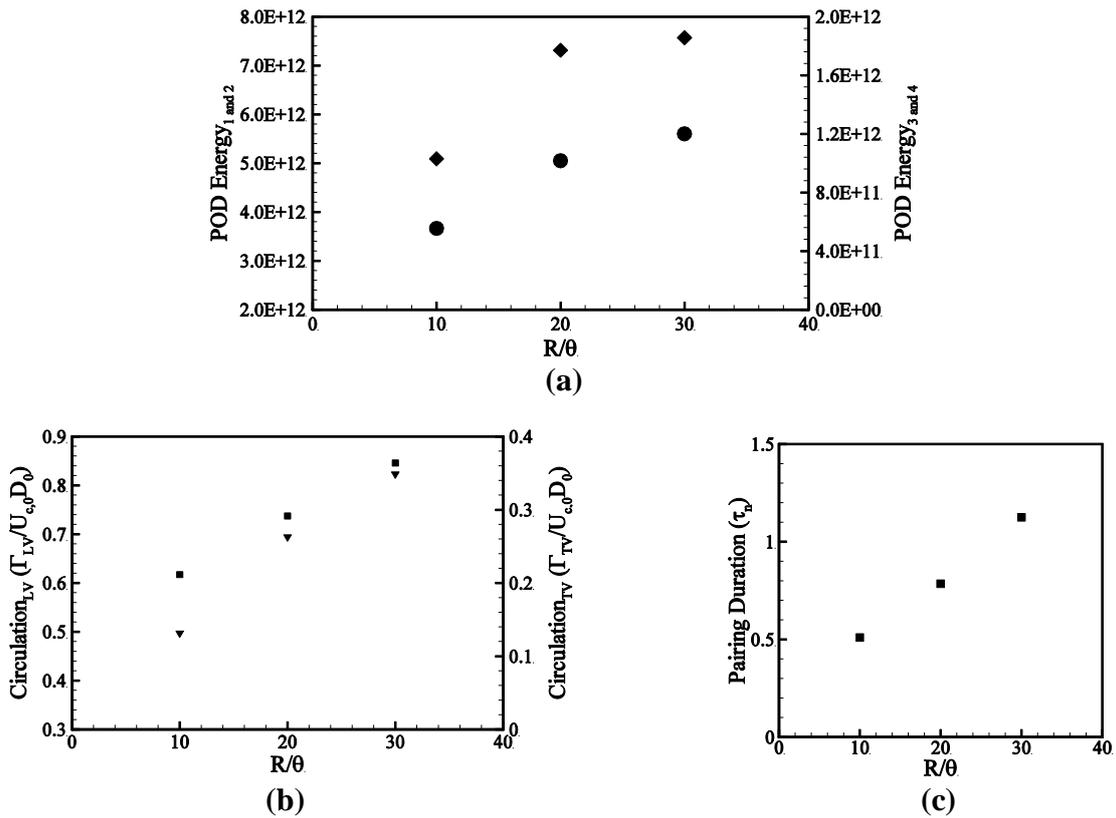

FIG. 23. Results of momentum thickness variation: (a) Variation of POD (kinetic energy of fluctuating velocity component), (b) Variation of circulation strength of leading and trailing vortex just at the start of vortex pairing, (c) Variation of the time duration of pairing (of leading and trailing vortices). Here, ▼ - circulation strength of trailing vortex, ■ - circulation strength of leading vortex, ♦ - combined POD energy for mode 1 and 2; ● - combined POD energy for mode 3 and 4.

*4.4. Variation of coflow temperature (St = 0.2)*

In this case, the jet in hot coflow is perturbed at the inlet at $St = 0.2$ to investigate the effect of hot coflow on the forced jet and the pairing process. As noted in Section 1, MILD combustion burners with hot coflow might affect the dynamics of the jet evolution owing to the temperature dependence of the viscosity and the changes in the density of the cold jet and hot coflow. Here, since both the cases show a similar effect of high coflow temperature on the coherent structures, the two cases of temperature ratios $(T_j/T_{cf}) = 0.5$ and 0.33 are discussed together for the brevity of the text. In both the cases, the coflow temperature is substantially higher than the cold jet core temperature. It makes temperature an important parameter to consider in the dynamics of coherent structures and the related analysis and calculations. Thus, a temperature based proper orthogonal decomposition is carried out on the two high-temperature cases, and the normalized eigenvalue is reported in Table 6. The temperature is coupled to the velocity according to the relation used in the previous studies [29-30]. According to this relation, if the temperature of the flow remains same, the POD energy (cumulative) matrix would be same as the velocity based energy matrix. Since the reference case has a constant temperature throughout its flowfield, a temperature-velocity based POD would have the same result as the velocity based POD. Hence, a velocity based POD on the reference case would be sufficient enough for the comparison with the cases of the jet in higher coflow temperature.

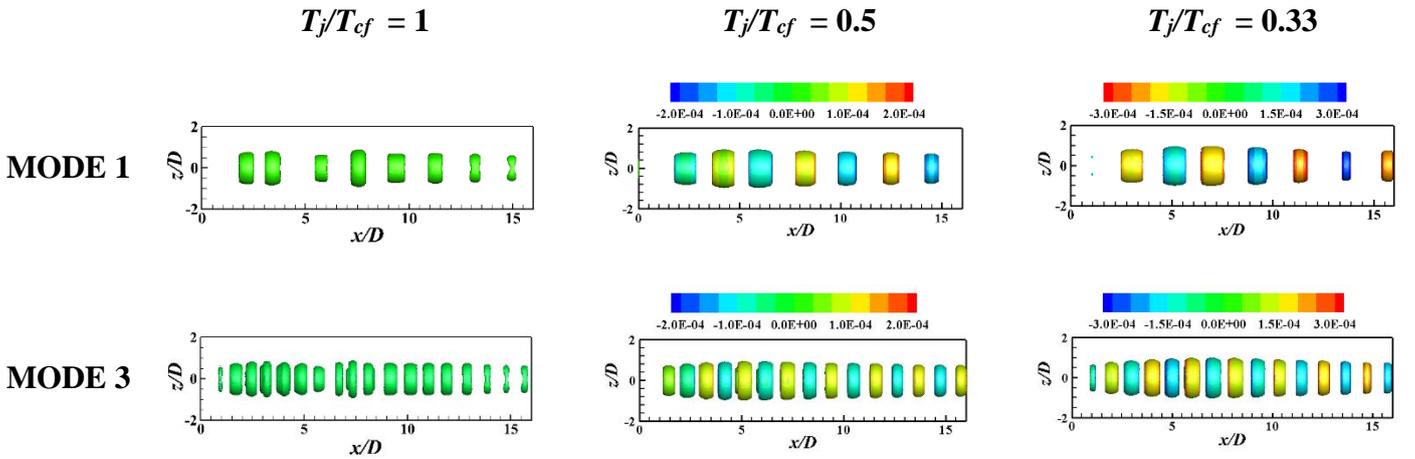

FIG. 24. Iso-surface of Q-criterion of Energy based POD modes for varying temperature ratio at $St = 0.2$.

The tabulated results (Table 6) and the iso-surface plots of Q-criterion of POD modes (Fig. 24) show the coupling between the consecutive modes, which is also visible in the plots of the temporal coefficients (not shown here) which are similar to the reference case. From Table 6, the energy of each mode is nearly same irrespective of the increase in coflow temperature and the trend of decrease in the energy across the POD modes for each case (of a particular temperature ratio) is same.

In the Fig. 24, the iso-surface plots of Q-criterion are colored with the temperature POD mode. The successive coupled modes 2 and 4 are not shown here because of the similarity in their structure to the coupled modes 1 and 3 respectively. The coherent structures obtained for each mode show alternate positive and negative values of eigenmodes for temperature. Usually, we perform POD on the fluctuating field, and the positive value (yellow and red colored vortices) is an indication of higher temperature while negative field (light/dark blue colored vortices) portrays the lower temperature. In Fig. 25(a), the vectors (related to velocity modes for $T_j/T_{cf} = 0.33$) heading downstream correspond to the region of vortices in the actual field while the vectors heading upstream correspond to braid region of the actual field. The blue colored region for temperature POD implies that the vortices are made of the colder jet-core fluid (because of the roll-up) while the hot fluid exists at the periphery of the braid region of the jet. The extension of the red contours near the jet center axis is indicative of the entrainment of the hot coflow in the jet.

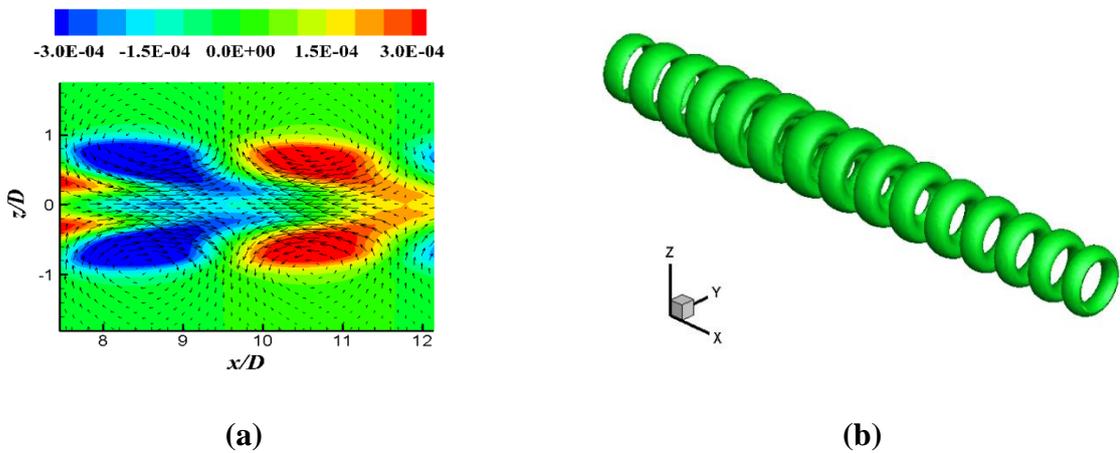

(a)            (b)

FIG. 25. (a) Vector plot of the velocities (POD modes) with the contour of POD mode of temperature in the background, (b) Iso-surface of Q- the criterion of mode 3 for jet in hot coflow with $T_j/T_{cf} = 0.5$, at St = 0.2.

The POD results and the series of snapshots indicate towards the absence of the pairing phenomenon between vortices. Unlike the reference case, the plot of Q-criterion (Fig. 25(b)) for the third mode does not show any vortex pairing. The structures in the third or fourth mode become stronger with an increase in the coflow temperature. However, the leading vortex is still generated by the normal process of a roll-up of the jet shear layer as explained in detail by Hussain [61], Martin & Meiburg [63]. The indented vortex structures in the middle of the jet near $x/D$ = 6 - 8 indicate two interacting vortices for $T_j/T_{cf}$ = 0.5. The lower temperature ratio $T_j/T_{cf}$ = 0.33 case also depicts the similar effect, where the indentation in the leading vortex fades away on moving further downstream.

*4.4.1. Circulation*

In case of $T_j/T_{cf} = 0.5$, from the circulation plots obtained by tracking a particular vortex shows that the vortex separates into two tiny vortices (Fig. 26). This small vortical ring starts separating out (disintegrating) while the formation of the leading vortex is still in process, with the same sign of vorticity. This tiny vortex ring maintains the coaxial concentric position concerning the primary (parent) vortex ring and gets dissipated while it remains in the same coaxial concentric position. The vorticity contours of the (parent) main vortex structure are shown in frames (in Fig. 26(a-e)) where the separation of the small vortex ring happens. Further, these instantaneous 2-dimensional frames with vorticity contours also show the dissipation of very weak trailing vortex just after being formed and before it undergoes pairing phenomenon (Fig. 26(f)). Here the first column shows the whole 2-dimensional domain with a highlighted vortex under study; the second column shows the vorticity contours of the given vortex, and the third column shows the given vortex at a higher minimum iso-vorticity contour value. The reason for having the third column as a part of our analysis is as follows:

The graphs are plotted to show the circulation (in m$^2$/sec) and area (in m$^2$) of the leading vortex (undergoing disintegration) against the time (in Fig. 27). The lowest value of iso-vorticity contours detected for the identification of vortices is 5.2x10$^5$ rad/s. With this as a minimum value of vorticity, the circulation and the area curve decreases uniformly after a small hump at the starting (as shown in Fig. 27(a-b)). While the continuous decrement is due to the dissipation, this hump occurs at a value of $\tau_n = 1.0$ for this case because of the similar change in the area of the vortex with time as shown in Fig. 27(b).

The increase in vortex area attributes to the disintegration of the vortex. Since the disintegration of vortices is not complete in this case, there are no significant results that can be found out from the circulation curve at this lowest value of iso-vorticity contour level depicted in Fig. 27(a-b).

| Mode | $T_j/T_{cf}$ | | |
|---|---|---|---|
| | 1 (ref.) | 0.5 | 0.33 |
| I | 43.09 | 39.66 | 40.09 |
| II | 41.92 | 38.79 | 37.93 |
| III | 5.97 | 6.59 | 6.72 |
| IV | 5.82 | 6.36 | 6.49 |
| V | 1.02 | 1.88 | 1.97 |
| VI | 1.01 | 1.85 | 1.93 |
| VII | 0.30 | 1.04 | 1.04 |
| VIII | 0.27 | 1.03 | 1.04 |

Table 6. Normalized eigenvalues of POD for the jet-in-hot coflow with varying temperature ratios ($T_j/T_{cf}$) at $St = 0.2$.

To quantify the amount of circulation being carried over to the two disintegrated vortices in the case of jet-in-hot-coflow, a similar analysis of calculation of circulation by vortex tracking is performed at 35 percent increased the value of lowest iso-vorticity contour level (viz. equal to $7.0 \times 10^5$ rad/s) as shown in Fig. 27(c-d). As noted in the earlier statements, a sudden dip in the curve is witnessed between $\tau_n = 1.0$ and $\tau_n = 1.05$ that suggests the point at which the separation (disintegration) completes at this higher value of lowest iso-vorticity contour. The amount of circulation that gets separated at 35% higher value of the minimum iso-vorticity contour is around 0.138 dimensionless circulation units. It may be a mere coincidence that the hump in the circulation curve for the minimum iso-vorticity contour (Fig. 27(a)) reaches its peak at $\tau_n = 1.0$. By assuming a value higher than the minimum iso-vorticity contour by some amount (viz. 35% here), the time instant of separation and the amount of separation of vorticity would vary.

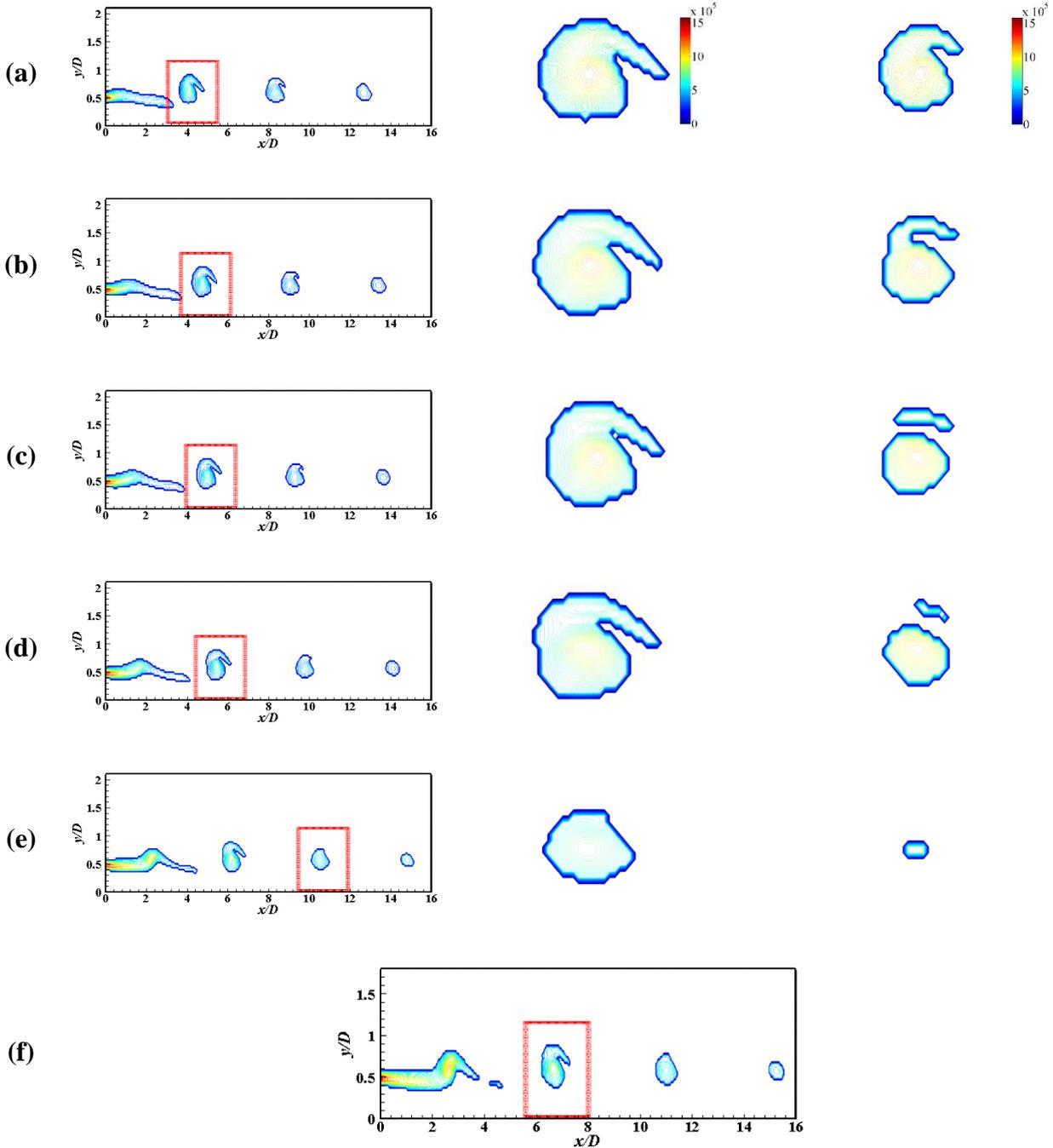

FIG. 26. Snapshots of vortex being studied for $T_j/T_{cf} = 0.5$ at $\tau_n =$ (a) 0.80 ( moments after separation from the shear layer at a lowest given value of the iso-vorticity contour), (b) 1.0 (just before separation of vortex into two as shown in third column), (c) 1.05 (after separation of vortex into two as shown in third column), (d) 1.2 ( after separation of vortex into two) , (e) 1.28 (after child vortex's complete dissipation as shown in third column), (f) Dissipation

of the weak trailing vortex just behind the (red highlighted) leading vortex.

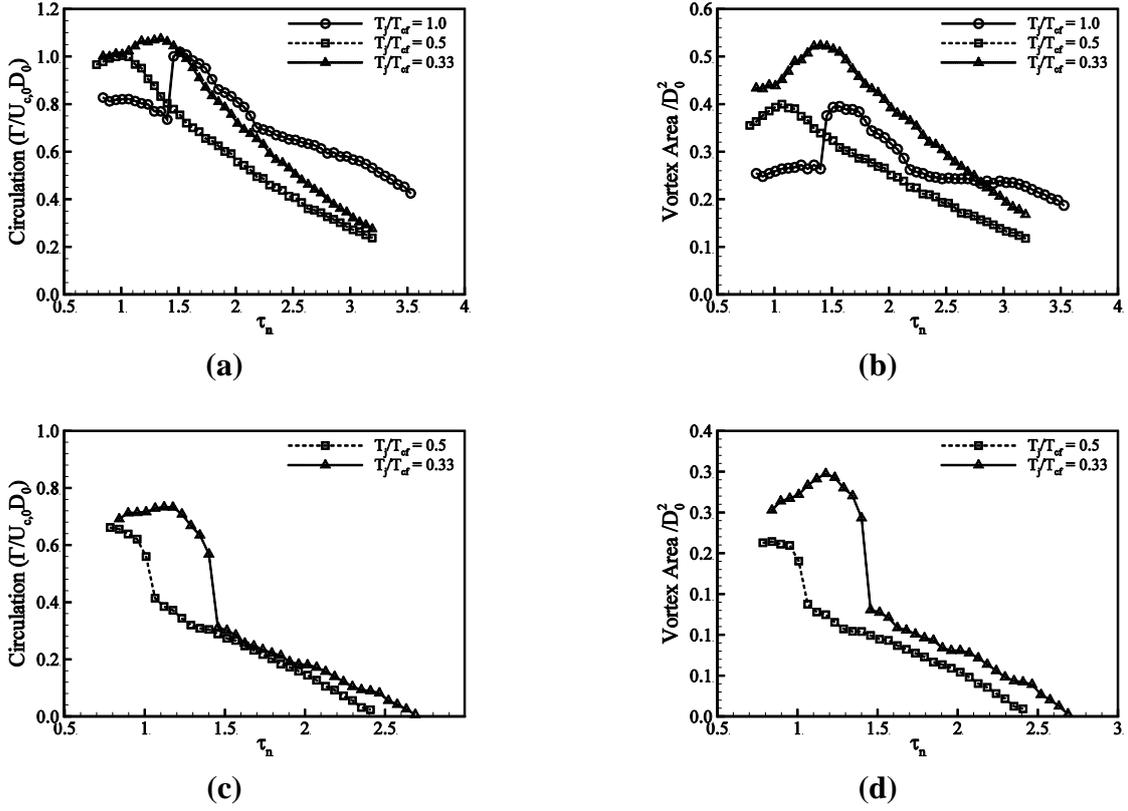

FIG. 27. Time variation of circulation and area of the leading vortex of jet-in-hot-coflow for different temperature ratios: (a-b) at the lowest iso-vorticity contour values, (c-d) at 35% higher than the lowest iso-vorticity contour values.

A typical value of 35% increment in the minimum iso-vorticity contour for both the hot coflow temperature cases leads to two things. On the one hand, at least two disintegrated vortices are distinguishable; whereas, on the other hand, the disintegrated smaller vortex remains for at least some downstream distance without undergoing its complete annihilation. Further, since the motive of this analysis at a higher value of the minimum iso-vorticity contour is to quantify the amount of separation in the leading vortex for higher temperature coflow cases, the relative values of separation (or disintegration) are provided here.

From the circulation plot (Fig. 27) for the case with $T_j/T_{cf} = 0.33$ at 35% higher value of the minimum iso-vorticity contour, the circulation dips to 0.261 non-dimensional circulation units at $\tau_n = 1.4$. The dip is nearly 1.9 times the dip observed in the circulation plot of the previous case having a lower coflow temperature. The dip also suggests that the circulation of the vortex, which undergoes incomplete separation from the leading vortex, increases 1.9 times when the coflow temperature increases 1.5 times. However, one has to perform the analysis at a value 35% higher than the minimum iso-vorticity contour level (used for the segregation of vortices from the shear layer and the vortices nearby). Owing to the similarity in the whole process which the coherent structure undergoes, we do not show the iso-vorticity contours for this lowest case of temperature ratio ($T_j/T_{cf} = 0.33$).

*4.4.2. Vorticity Budget Equation*

Rather, to further study the cause of vortex separation and the contribution of the hot coflow to it, we have analyzed individual terms in vortex budget equation (Eq. 6) for only one of the cases - $T_j/T_{cf} = 0.33$. The term-by-term analysis of vorticity transport Eq. (6) sheds light on the vortex structures appearing in the increased temperature case. The contour plot (Fig. 28 (a-d)) suggests that the stretch term contributes the most followed by baroclinic, dilatation and diffusion terms respectively. The comparison of the terms for hot coflow to the cold coflow base case suggests that the stretch term of hot coflow is one of the major causes for bifurcation of the vortex structure (Fig. 28(a) and Fig. 29(a)). The third term of the equation, the baroclinic term has an almost equivalent effect on the vortex evolution in hot coflow case, but it is insignificant in the cold coflow case (almost 100 times smaller) (Fig. 28(c) and Fig. 29(c)). While this term has almost no effect in cold flow case, it is one of the reasons that can be the cause for bifurcation in hot coflow case. Similarly, second term, i.e., the dilatation is smaller than stretch term by order of magnitude in cold coflow case but has a dominating effect in hot coflow case (Fig. 28(b) and Fig. 29(b)). The fourth term is the dissipation, which plays a more dominating role in cold coflow case compared to hot coflow in vorticity transport equation (Fig. 28(d) and Fig. 29(d)).

The negative (dark blue) and positive (red) regions on the side of the indentation in the vorticity contour represent the contribution of the baroclinic term as depicted in Fig. 28(b). The pattern is somewhat similar to the cold coflow jet (Fig. 29(b)) except at the vortex end that faces the inlet (that is, negative (blue) in case of hot coflow and positive (red) in case of cold coflow). The term is significant at places where the pressure gradient is perpendicular to the density gradient. Since in cold coflow cases the variation of density is negligible, the contribution to the vorticity by the baroclinic term is insignificant.

The positive contribution on the top of the indentation helps the increment of the separated structure while the negative contribution below the indentation prevents the increment of the primary (leading) vortex structure. The counteract on either side of the indentation helps to sustain the separated structure of the region contributed positively by the baroclinic term and prevent its early dissipation. From the plot of the total rate of change in vorticity (see Fig. 28 (i)) it is observed that there is an increment in the front region of the two separating vortical structures (red) and decrement in the region near the indentation of the vorticity contours (dark blue). This negative projection (along with the indentation of vorticity contours) in hot coflow case is absent in the cold coflow jet.

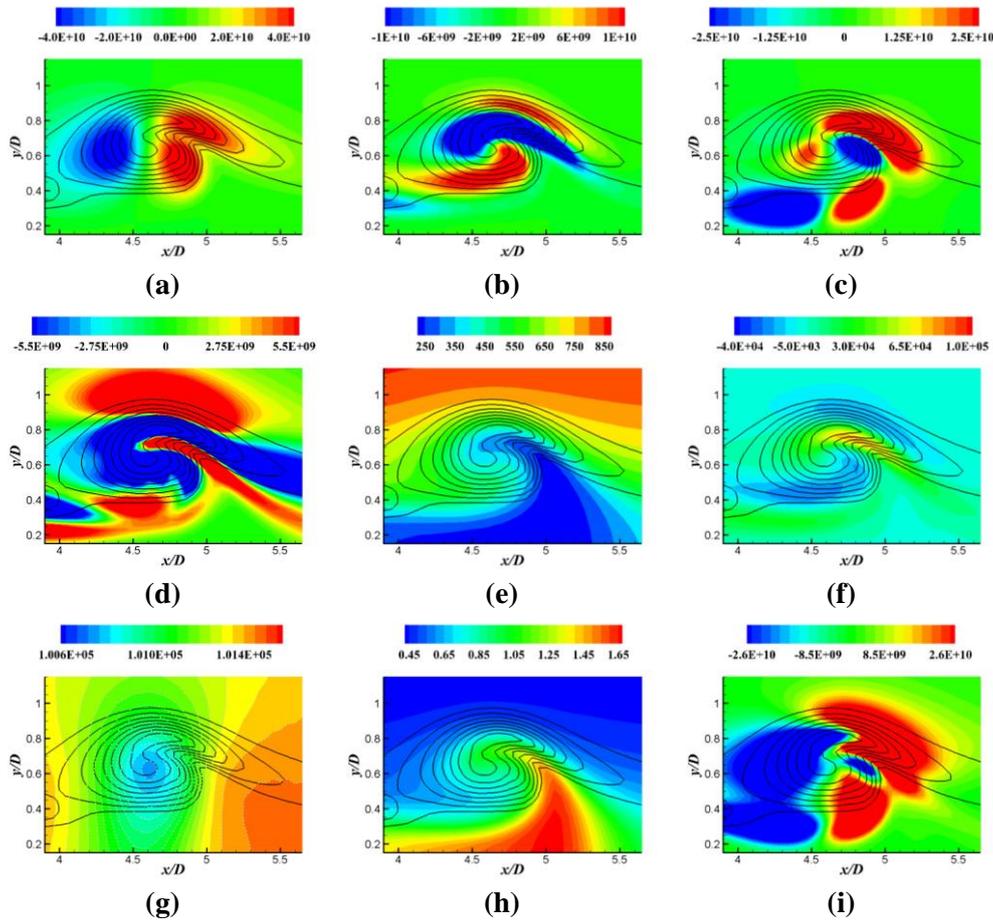

FIG. 28. Vorticity budget analysis (Eq. 6) for $T_j/T_{cf} = 0.33$ at $St = 0.2$: (a) Stretch term, (b) Baroclinic term, (c) Dilatation term, (d) Diffusion term. Contour plots of (e) Temperature, (f) Divergence of Velocity, (g) Pressure, (h) Density, (i) Total Rate of change of Vorticity plot.

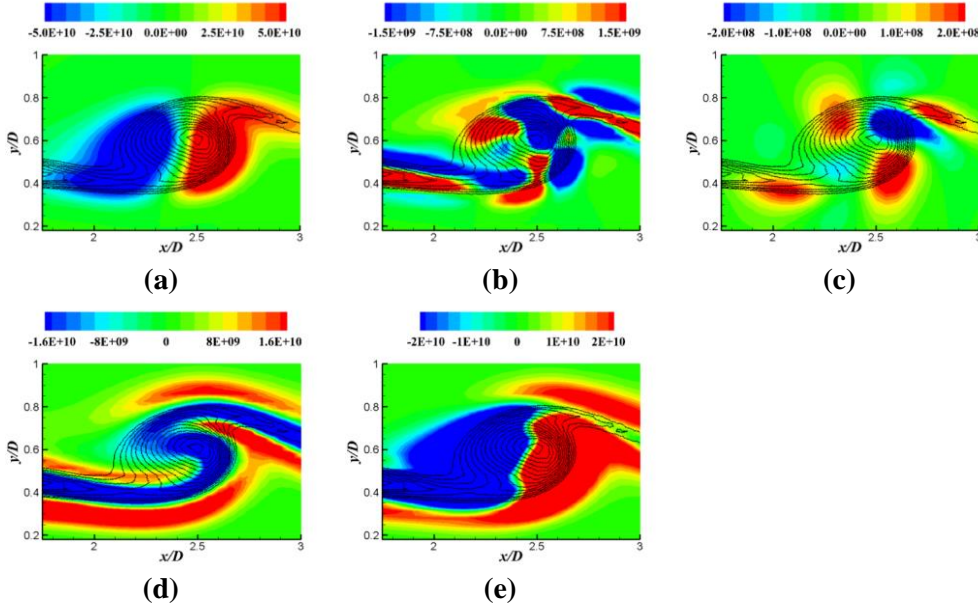

FIG. 29. Vorticity budget analysis (Eq. 6) for $T_j/T_{cf} = 1.0$ (cold coflow) at $St = 0.2$: (a) Stretch term, (b) Baroclinic term, (c) Dilatation term, (d) Diffusion term, (e) plot of the total rate of change of vorticity.

To further explain the baroclinic effect, a schematic along with the simulated results is shown in Fig. 30. In Fig. 30(a), the pressure varies from left to right, and density differs for the two fluids (with denser fluid at the bottom). This pressure-density variation causes the lighter fluid to gain higher velocity ($u_1$) due to its lesser inertia of rest as compared to the velocity ($u_2$) of denser fluid, hence inducing a clockwise (negative) vorticity for the case shown (in Fig. 30(a)). A simplified cartoon image (Fig. 30(b)) can explain the effect of the baroclinic term on the vorticity (marked in red) of the jet in hot coflow (with variable density. The image helps in identifying the region of anti-clockwise (positive) vortices contributing to the vortex growth and vice versa. Here the red colored arrows show the direction of rotation of the vortices.

The dark blue region along the indentation in the vorticity contours represents the negative contribution of the gas expansion due to the expansion of the cold fluid in the given region (Fig. 28(b)). One can easily understand this by attaching a small control volume to the core of the vortex (Fig. 30(c)) which moves along with the vortex (at same local convection speed). Through the front end of the control volume cold denser fluid enters (as shown in Fig. 30(c)), circulates within it and mixes with hot coflow fluid and exits from other sides of the control volume. During its residence within the control volume, it absorbs the heat from the surrounding due to heat diffusion and hence its temperature rises and fluid results in positive expansion. This expansion decreases the strength of vortex (circulating fluid) contributing to negative vorticity. The temperature and the density contours further confirm the same (Fig. 28(e) and (h)).

It can be said that these two terms together have predominantly caused the decrement of net vorticity in the middle of the leading vortex (small dark blue patch in Fig. 28 (i)) leading to the increased indentation of vorticity contours. Hence, the temperature variation across the jet shear layer is a 'direct cause' of tearing/separation of the leading vortex here. The motivation towards analyzing the effect of Reynolds number and coflow velocity variation on the vortex development paves the way for the next two following sections. Since the pairing and the circulation of vortices are being the focus of the following sections, we have only investigated the circulation plots for the respective variations.

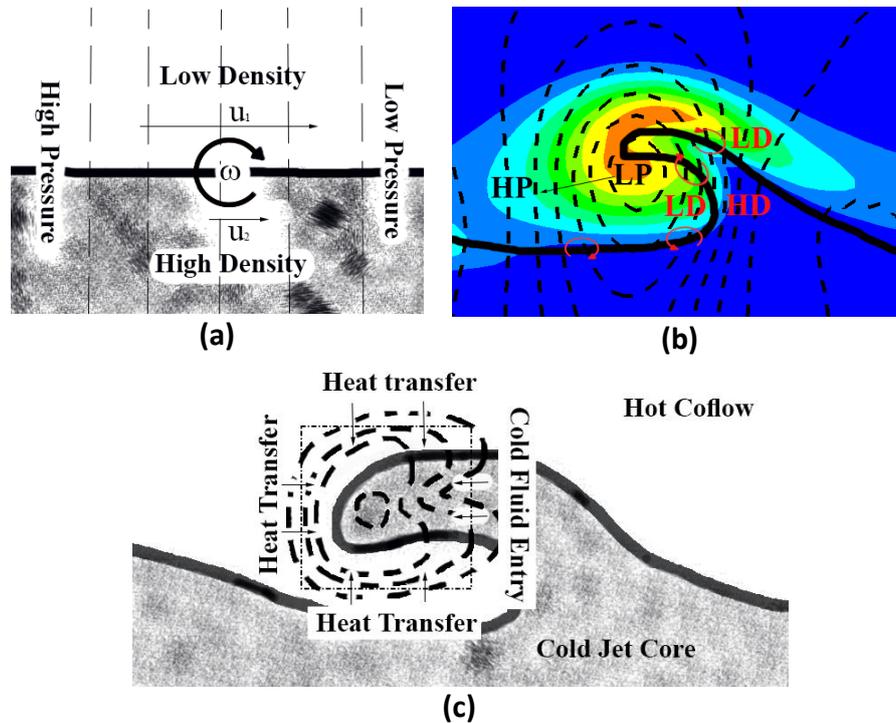

FIG. 30. (a) Pictorial representation of the baroclinic induced vorticity. Here $u_1$, $u_2$ - velocity of low density and high-density fluids and $\omega$ is the clockwise (negative) vorticity induced. (b) Baroclinic induced vorticity production and dissipation due to positive and negative vortices (red arrows). Here ─ ─ ─ denotes pressure contours, ─── demarcates the boundary between high-density cold jet (below) and low-density hot coflow (top). Abbreviations: HP - High Pressure, LP - Low Pressure, HD - High Density, LD - Low Density. (c) Expansion of cold jet fluid because of the hot coflow and its effect on the vorticity during the roll-up process. Here ─ ─ ─ denotes vorticity, ─ · · ─ · · ─ represents the boundary of the moving control volume and ─── demarcates the boundary between cold jet (below) and hot coflow (top). Arrows indicate the direction of heat transfer or entry of the cold fluid at their respective places into control volume.

As has been noticed above, the disintegration of the leading vortex ring takes place, and the proportion of the separating vortex's circulation increases as the coflow temperature increases. Particularly in the present study, the data of circulation of the separating vortex increases 1.9 times by increasing the coflow temperature by 1.5 times. The analysis of the vorticity budget equation yields the reason of separation (disintegration) of the vortices, which primarily happens due to the combined effect of gas expansion and baroclinic terms in causing the attenuation of the vortex in the middle while strengthening on either side. The Kelvin-Helmholtz instability induced mixing of cold jet fluid with hot coflow fluid has been found to be responsible for the disintegration of vortices.

*4.5. Variation of coflow velocity*

In this section, we have reported the variation in coflow velocity. Altering the coflow strength would affect the velocity gradient and thus, the vortex structures. The aim of these set of simulations is to determine the effect of coflow on the relative strength of the leading and trailing vortex and the time instants of their formation. Two cases - one with low coflow velocity (5 % of jet core velocity ($U_{jet}$)) and another with higher coflow velocity (15 % of $U_{jet}$) than the reference case (10 % of $U_{jet}$), are discussed. The tracking of leading vortices in the jet core allows calculating their circulation which is then plotted against the non-dimensional time (as discussed earlier) in Fig. 31. The minimum detectable iso-vorticity contour value found for three cases are - 5.2 x $10^5$ m/s, 5.6 x $10^5$ m/s and 6.0 x $10^5$ m/s (in the order of increasing coflow velocity).

The circulation for lower coflow (5 % of $U_{jet}$) and the reference coflow velocity (10 % of $U_{jet}$) are similar but different than the case with higher coflow velocity (15 % of $U_{jet}$). In higher coflow case, there exists a small and sudden jump after a considerable drop in the circulation plot, followed by a gradual rise. The decrement in the circulation is because of the dissipation of vortices while the sudden jump and the gradual increment are due to the pairing of the leading and the trailing vortices.

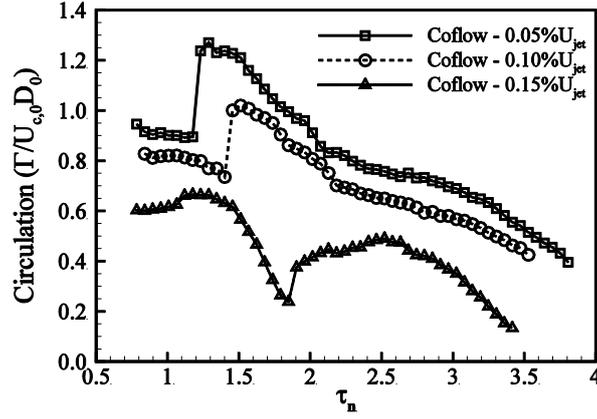

FIG. 31. Circulation with non-dimensional time for varying coflow velocity at $St = 0.2$.

Two points are important in this case: firstly, the lower starting value of circulation of the leading vortex compared to others; secondly, the maximum circulation achieved after the sudden jump (vortex pairing) is less than the starting circulation value. The former can be explained by the less velocity difference between jet and coflow while the later indicates weaker participating trailing jet vortex. Considering all three cases, the sudden jump in the circulation of leading vortex decreases as the coflow velocity increases. Since velocity difference decreases with increasing coflow velocity, the circulation of trailing vortex decreases which leads to a lower jump in the circulation. The circulation jump with the increasing coflow velocities - 5%, 10% and 15% of $U_{jet}$ - are found to be 0.346, 0.261 and 0.135 non-dimensional circulation units, respectively. The reference case has also depicted similar curve, and the circulation trend for the case with lower coflow velocity (5 % of $U_{jet}$) can be explained similar to an earlier discussion.

Another key observation here is the delayed vortex pairing as the coflow velocity increases. As discussed by Martin & Meiburg [63], global and local induction plays a vital role in the dynamics of jet and hence affects the vortex pairing as well. As coflow velocity increases, the local induction effects of both the vortices, i.e., leading and trailing, decreases. These weak induction forces allow more time for the vortices to travel further downstream before they undergo vortex pairing, which explains the delayed sudden jump with increasing coflow velocity. The increase in the non-dimensional time instant ($\tau$) at which the vortex pairing starts with the increasing coflow velocity - 5%, 10% and 15% of $U_{jet}$ - is found to be - 1.18, 1.4 and 1.85, respectively.

*4.6. Variation of Reynolds number*

In this section, the discussion is related to the variation in jet Reynolds number (*Re*). We have considered two cases with lower *Re* = 500 and higher *Re* = 1500 along with the reference case of *Re* = 1000. The vortex evolution is studied, and the circulation is plotted for each case as shown in Fig. 32. The lowest detectable iso-vorticity contour for the three cases with increasing *Re* are - 2.5 x $10^5$ m/s, 5.6 x $10^5$ m/s and 8.2 x $10^5$ m/s respectively.

The sudden jump in the circulation curve for the cases with *Re* = 1000 (reference case) and *Re* = 1500 show vortex pairing as against the case with *Re* = 500. As observed from the study of the snapshots of the case with *Re* = 500, the generated weak trailing vortices dissipate soon after its production before getting involved in vortex pairing. The low values of the circulation strength and the weak mutual induction effects by the leading and the trailing vortices, in this case, are insufficient to cause vortex pairing before the vortices dissipate. Also, Fig. 32 confirms the moment (at $\tau_n = 0.84$) right after the separation from the shear layer at a lowest given value of the iso-vorticity contour. The trailing vortex is found to annihilate in between $\tau_n = 1.11 - 1.14$. Finally, at $\tau_n = 3.15$ is the time instant just before the leading vortex dissipates. Davies and Baxter [64] reported the delayed development of Kelvin-Helmholtz instability and the vortex pairing, as the viscous damping strengthened in low Reynolds number. In the current work, increase in the viscous damping is the primary cause of dissipation of the weak trailing vortex even before pairing. Hence the vortex pairing is not observed in this low Reynolds number case.

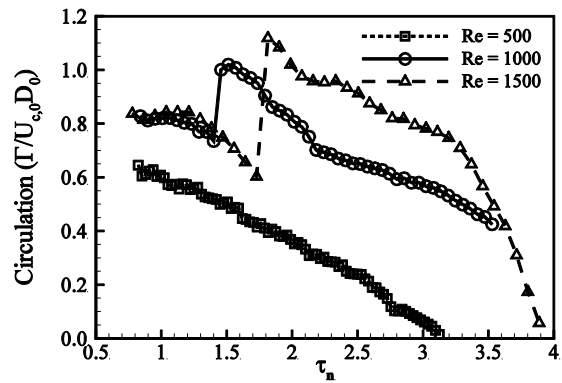

FIG. 32. Circulation with non-dimensional time for varying Reynolds number at *St* = 0.2.

In higher Reynolds number ($Re = 1500$) case, the curve (in Fig. 32) exhibits strong vortex pairing but also a sudden drop is visible after non-dimensional time instant $\tau_n = 3.2$. This drop indicates the quick dissipation of the vortices as the jet undergoes a transition to turbulence. The transition of the jet into turbulence is visible in Fig. 32(d). The sudden jump in the circulation curve (in Fig. 32) is primarily because of strong vortex pairing, i.e., 0.513 non-dimensional circulation units (almost twice the reference case) but is delayed by 0.3 non-dimensional time instants (at $\tau_n = 1.7$ as compared to 1.4 of the reference case). At the time step $\tau_n = 3.9$, the leading vortex completely dissipates. The dissipation of the leading vortices is present in both the lower and higher $Re$ but not in the case of $Re = 1000$ (reference case). It might be because of the higher viscous effects at lower $Re = 500$ whereas it is the onset of turbulence, which the (paired structure) vortex ring annihilates entirely at the higher $Re = 1500$.

The absolute magnitude of circulation strength of the vortices is higher at $Re = 1500$ as compared to $Re = 1000$ because of the strong velocity gradient between the jet core and the coflow (since circulation is non-dimensionalized, it is not visible in this Fig. 32). The late vortex pairing observed in the case of $Re = 1500$ is contradictory to the results obtained by past researchers [1], [3], [64]. This is because, in the previous works, the vortex shedding frequency was same as that of the exciting frequency applied at the inlet in Reynolds number regime where the jet is stable and axisymmetric. In past work, the vortices generated were the direct effect of the perturbation given at the inlet but in this work, two vortices - leading and trailing vortices – are produced for a single perturbation given at the jet inlet. Since the time period of the vortices is very small in $Re = 1500$ as compared to $Re = 1000$, the delay in temporal domain is visible in the vortex pairing for a jet with $Re = 1500$. (In the absolute sense, the time instant of vortex pairing in a jet with $Re = 1500$ is before than that of the jet with $Re = 1000$)

## 5. Discussion

Low Reynolds number forced (varicose) jet-in-a-coflow is studied computationally for a range of varying parameters to understand the dynamic evolution of coherent structures. The parametric variation includes forced (varicose) perturbation frequency, coflow temperature, momentum thickness, and turbulence intensity at the inlet and Reynolds number. One of the critical observations, as also observed by Michalke & Hermann [50], is the stability achieved by jet in the presence of coflow. The evolution process of coherent structures in the current scenario is different from that of a pure jet without coflow. On comparing the present work with that of Gohil *et al.* [14], the results of perturbation of jet at $St = 0.2$ and $0.3$ appear to be similar to the jet with and without coflow, as noted and mentioned earlier in Section 4.1. However, the results obtained for the jet in coflow with $St = 0.5, 0.6$ show vortex pairing in the Q-criterion iso-surface contours as well as in the FFT results. These results are well in agreement to the oscilloscope traces of the jets with $St = 0.7 - 1.0$ and the FFT results reported for a jet with $St = 0.85$, by Zaman and Hussain [3]. In the work of Gohil *et al.* [14], vortex pairing is not visible in the iso-surface contours of Q-criterion for the pure jet [$St = 0.6$, $Re = 1000$]. In their work, the jet with $St = 0.6$ show increased instability on the vortex rings and development of streamwise vortices. Since no FFT or any other detailed analysis is carried out in their case, it is difficult to conclude about the absence of vortex pairing. The absence of streamwise vortices and instabilities in the vortex structures, for the current case of the jet in coflow with $St = 0.6$ further indicate the stability effects of coflow.

Although the two cases of $St = 0.2$ and $0.6$ show vortex pairing, the FFT results are found to be significantly different. For $St = 0.2$ case, the fundamental peak is absent at the inlet ($x/D = 0$ in Fig. 7). The appearance of the fundamental peak at $x/D = 2$ is due to the formation of the trailing vortices. Upon completion of the process of vortex pairing, the fundamental peak disappears (as observed at $x/D = 12$ in Fig. 7). The observed fact is similar to the finding (FFT result) of a jet without a coflow case at $St = 0.2$ (Gohil *et al.* [14]). In contrast, the jet in coflow at $St = 0.6$ shows a fundamental peak at the inlet which vanishes after the vortex pairing completes (observed at $x/D = 10, 13$ in Fig.6). Vortex pairing being a common phenomenon taking place in both the cases, the sub-harmonic frequency is also noticed to grow stronger as one moves in the downstream direction of the flow domain, along with the jet centerline axis.

As observed from the investigation of the cases of turbulent jet-in-coflow, the energy based POD analysis is not able to accurately predict the mode with the vortex pairing phenomenon, the frequency based DMD analysis is found to be a better marker of the coherent structure visualization (also discussed in Section 4.2.2). The coherent structures based on the temporal orthogonality (in case of DMD) not only help in conveniently locating the mode related to vortex pairing and the paired structures (which has characteristic sub-harmonic frequency) but also the vortex structures related to the fundamental frequency.

Further, the analysis of the effect of varying momentum thickness ratio on the coherent structure formation is reported in Section 4.3. Some conclusions that one can draw from the examination of the results discussed earlier (in later-half of Section 4.3) – firstly, total circulation is increased with increase in momentum thickness ratio. This is apparent from the higher velocity gradient in case of the increased momentum thickness ratio. Secondly, circulation capacity of the leading vortices increases with the increase in momentum thickness ratio. Thirdly, increase in circulation of trailing vortices (with the increase in momentum thickness ratio) tells that the rate of increase of total circulation (leading + trailing) is higher than the rate of increase of circulation capacity of leading vortices.

From this, it can be inferred that the relative amount of circulation as well as the energy of the leading vortex decrease with the increase in $R/\theta$. According to the study of Gharib *et al.* [12], Mohsini *et al.* [13] and Dabiri *et al.* [65], there is a maximum limit of circulation and energy of the vortices formed for a jet beyond which the additional circulation causes the formation of trailing vortices. Similarly, in the present scenario, it can be said that the maximum limit of circulation/ energy of a leading vortex increase with an increase in $R/\theta$ but the supply of circulation/ energy at the inlet (through shear layer) is even higher than this limit of the leading vortices. This leads to the formation of the trailing vortex with increased circulation and energy. Apart from the pairing of the leading and the trailing vortices, an unusually gentle, positive slope of circulation in the case of $R/\theta = 10$ (occurs after the vortex pairing event in Fig. 22(a), (c) and (d)) indicates the addition of vorticity to the already-paired vortex ring structure. It has been shown from the instantaneous vorticity contour plots (in Fig. 19) that a second vortex pairing is the cause of slight increment in the circulation. Circulation plot also shows the tearing of the vortex at a downstream distance of $x/D = 13$ for the $R/\theta = 10$ case (Section 4.3.1). However, the POD modes do not capture the structures corresponding to the second vortex pairing and the tearing process.

As observed in Section 4.3 (Fig. 23(c)), the increase in the vortex pairing time duration can be related to the decrease in the relative difference of circulation/energy of the trailing and the leading vortices [66]. The leap-frog motion is noticed in the current data between the leading and the trailing vortices which is similar to the results present in the literature [3, 67]. Hence, this kind of motion should go for a longer duration for vortices of the same size before their complete mergence.

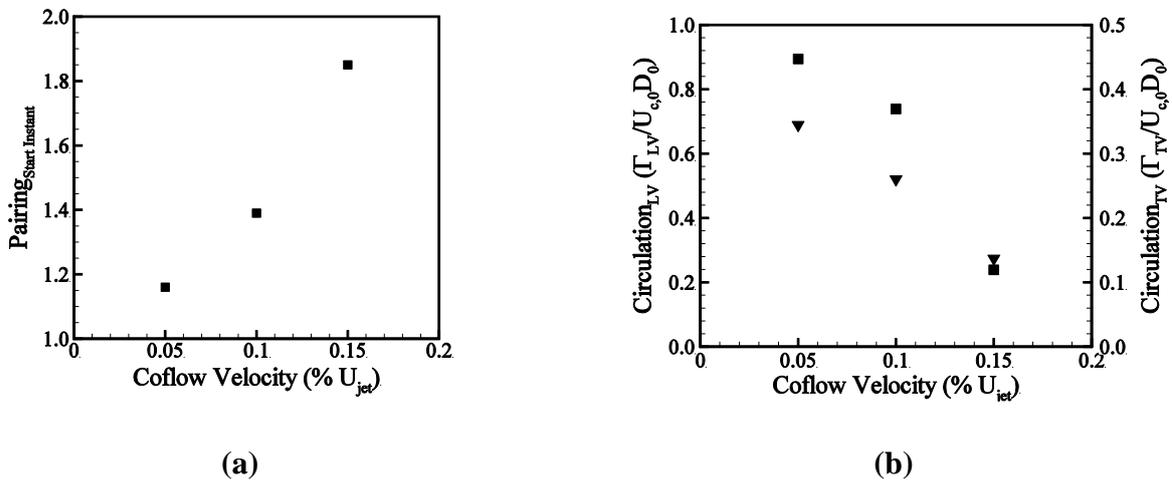

FIG. 33. Results of co-flow variation: (a) Variation of starting instant (time), (b) Variation of circulation strength of leading and trailing vortex just at the start of vortex pairing. Here, ▼ - circulation strength of trailing vortex; ■ - circulation strength of leading vortex.

The analysis reported in Section 4.4 focuses on the effect of coflow temperature on vortex structures. The vortex pairing is absent in third or the subsequent modes. Further, the reconstruction of these modes (not shown here) shows smaller trailing vortices which disappear before vortex pairing, which is also visible in the iso-vorticity contours in Fig. 26. This fact corroborates the higher stability of the jet at an increased coflow temperature as compared to the reference case (cold coflow). The higher magnitude of circulation has been observed for the vortex rings in lower temperature ratio cases (see Fig. 27(c)). Since the circulation is dependent on the vorticity magnitude at a given point and the area,

Fig. 27(b) exhibits the vortex area with respect to time. The trend observed for both – the circulation and the area of the vortex – is similar. Hence it can be said that the circulation of the vortex is dependent on the area. The increase in circulation capacity of the leading vortices with a decrease in temperature ratio is associated with the stability achieved by both vortical structures and jet shear layer with an increase in the temperature of the shear layer. Demetriades *et al*. [68] had conducted an experimental study of laminar-turbulent transition in free shear layers and had shown that the transition Reynolds number is proportional to the dividing streamline temperature. The dividing streamline temperature has the effect of temperature from both the sides of the free shear layer, i.e., inner jet core and outer coflow – for the present case. Hence a hot coflow surrounding a cold jet, in the present case, would increase the dividing streamline temperature of the jet free shear layer, thereby leading to increased stability and the transitional Reynolds number. Consequently, the jet circulation capacity of the vortical structures also increases which is the reason for having very weak trailing vortices in lower temperature ratio cases of jet-in-coflow. The Fig. 26(f) shows weak, smaller trailing vortex behind the leading vortex enclosed by a red colored rectangle. As noted above, the pairing process is also not observed as the weak trailing vortices get dissipated soon after they form behind the leading vortices.

It can be said while looking at Fig. 31 that increasing the coflow velocity to 15% of $U_{jet}$ at a particular *Re* has decreased the strength of leading and trailing vortex just before they undergo vortex pairing. Moreover, the pairing delays, and it occurs after a considerable drop in circulation strength of the vortices due to viscous dissipation (Fig. 31). When the coflow velocity decreases to 5% of $U_{jet}$ compared to the reference case (10% of $U_{jet}$), vortex pairing occurs at an earlier time instant (Fig. 31), and the strength of trailing and leading vortices (measured just at the instant of pairing) enhance. Fig. 33 shows the trends of the starting time instant of vortex pairing and strength of the leading and trailing vortices for different coflow. However, we have not shown the vortex pairing duration for this case as the case of 15% coflow undergoes immediate merging of the vortices.

In case of increased Reynolds number (*Re* = 1500), the jump in circulation strength of the leading vortex at the start of vortex pairing is two folds as compared to the reference case of *Re* = 1000 (Fig.

32). Although the absolute value of the circulation of the leading vortex remains constant, the relative value with respect to the trailing vortex decreases. Additionally, the delayed vortex pairing observed in Fig. 32, is followed by a sudden drop in the circulation strength of the paired structure because of transition to turbulence (as evident from the shape and size of the paired structure in various snapshots, not shown here). Whereas, on decreasing the Reynolds number to 500, no vortex pairing exists due to weak local induction effect of the leading and the trailing vortices on each other and an early dissipation of the trailing vortex before pairing.

## 6. Conclusion

In this work, DNS study has been carried out for the forced jets in coflow under different conditions and analyzed using various post-processing tools like fast Fourier transform (FFT), Proper Orthogonal Decomposition (POD), Dynamic Mode Decomposition (DMD) and the vortex tracking method. The jet spreading obtained at different frequencies of forced perturbations is maximum for Strouhal number 0.3 which is in agreement with the previous work on a jet without coflow. The rest of the study focuses on the jet-in-coflow forced at Strouhal number ($St$) of 0.2 for its peculiarity of involving the trailing and the leading vortices in the pairing process.

Some of the notable results of the POD on the jet-in-cold-coflow are the coupling of the consecutive modes having a similar amount of energy. The POD can distinguish between the leading and the trailing vortices which provides some useful information on the (fluctuating) kinetic energy of the concerned structures. Further, the jet-in-coflow perturbed at different inlet turbulent intensities has uncoupled POD modes showing smaller turbulent structures which possess more dominating energy than the more significant coherent (vortical) structures. The POD is unable to capture not-so dominating pairing phenomenon due to its spatial orthogonality only. However, the DMD of the flow field, based on frequency, can predict the vortex structures concerned to the region before and after the pairing. This also shows that pairing phenomenon is not as dominant energy-wise as it is in frequency compared to the turbulent structures formed elsewhere.

While varying the coflow temperature, the circulation capacity increases for the leading vortices due to increased stability of the shear layer, whereas it decreases for the trailing vortices as the temperature ratio decreases in case of jet-in-hot-coflow. This leads to the early dissipation of the trailing vortices even before it undergoes pairing. For the jet in hot coflow case, the disintegration of the leading vortices occurs near the jet inlet. It is not the variation in circulation capacity but the mixing of the cold jet core with hot coflow fluid (in the K-H induced instabilities), which is found to be responsible for the initiation of vortex disintegration.

Further, the simulations on cases with varying momentum thickness show that the ratio of the energy of the leading vortices to that of the trailing vortices decreases when the non-dimensional momentum thickness ratio of the jet in-coflow increases. The amount of the turbulent kinetic energy carried by the leading and the trailing vortices increase with an increase in the non-dimensional momentum thickness ratio. Similarly, the circulation of leading as well as trailing vortices along with the vortex pairing duration is found to increase as the non-dimensional momentum thickness ratio increases.

With the increase in the coflow velocity, weaker and delayed vortex pairing takes place. The strength of the leading and the trailing vortices also decrease, but with the increase in Reynolds number, the vortex pairing strengthens. However, in the case of low Reynolds number ($Re = 500$), the weak trailing vortices dissipate even before the start of vortex pairing.

**Acknowledgement**

The authors would like to acknowledge the IITK computer center (www.iitk.ac.in/cc) for providing the resources for performing the computation work, data analysis, and article preparation. Also, we would like to thank Prof. Debopam Das of AE, IITK for his valuable inputs regarding the manuscript preparation.

[3] Zaman, K. B. M. Q., & Hussain, A. K. M. F. (1980). Vortex pairing in a circular jet under controlled excitation. Part 1. General jet response. *Journal of fluid mechanics*, *101*(3), 449-491.

[4] Ho, C. M., & Huang, L. S. (1982). Subharmonics and vortex merging in mixing layers. *Journal of Fluid Mechanics*, *119*, 443-473.

[5] Ho, C. M. (1982). Local and global dynamics of free shear layers. In *Numerical and Physical Aspects of Aerodynamic Flows* (pp. 521-533). Springer, Berlin, Heidelberg.

[6] Ho, C. M., & Nosseir, N. S. (1981). Dynamics of an impinging jet. Part 1. The feedback phenomenon. *Journal of Fluid Mechanics*, *105*, 119-142.

[7] Laufer, J., & Monkewitz, P. (1980). On turbulent jet flows-a new perspective. In *6th Aeroacoustics Conference* (p. 962).

[8] Kibens, V. (1980). Discrete noise spectrum generated by acoustically excited jet. *AIAA Journal*, *18*(4), 434-441.

[9] Kelly, R. E. (1967). On the stability of an inviscid shear layer which is periodic in space and time. *Journal of Fluid Mechanics*, *27*(4), 657-689.

[10] Corcos, G. M., & Sherman, F. S. (1976). Vorticity concentration and the dynamics of unstable free shear layers. *Journal of Fluid Mechanics*, *73*(2), 241-264.

[11] Riley, J. J., & Metcalfe, R. W. (1980, January). Direct numerical simulation of a perturbed, turbulent mixing layer. In *18th Aerospace Sciences Meeting* (p. 274).

[12] Gharib, M., Rambod, E., & Shariff, K. (1998). A universal time scale for vortex ring formation. *Journal of Fluid Mechanics*, *360*, 121-140.

[13] Mohseni, K., & Gharib, M. (1998). A model for universal time scale of vortex ring formation. *Physics of Fluids*, *10*(10), 2436-2438.

[14] Gohil, T. B., Saha, A. K., & Muralidhar, K. (2013). Direct numerical simulation of forced circular jets: effect of varicose perturbation. *International Journal of Heat and Fluid Flow*, *44*, 524-541.

[15] De, A., & Dongre, A. (2015). Assessment of turbulence-chemistry interaction models in MILD combustion regime. *Flow, Turbulence and Combustion*, *94*(2), 439-478.

[16] Bhaya, R., De, A., & Yadav, R. (2014). Large eddy simulation of MILD combustion using PDF-based turbulence–chemistry interaction models. *Combustion Science and Technology*, *186*(9), 1138-1165.

[17] Dongre, A., De, A., & Yadav, R. (2014). Numerical investigation of MILD combustion using multi-environment Eulerian probability density function modeling. *International Journal of Spray and Combustion Dynamics*, *6*(4), 357-386.

[18] Huang, X., Tummers, M. J., & Roekaerts, D. J. E. M. (2017). Experimental and numerical study of MILD combustion in a lab-scale furnace. *Energy Procedia*, *120*, 395-402.

[19] Xing, F., Kumar, A., Huang, Y., Chan, S., Ruan, C., Gu, S., & Fan, X. (2017). Flameless combustion with liquid fuel: A review focusing on fundamentals and gas turbine application. *Applied Energy*, *193*, 28-51.

[20] Dobler, W., & Brandenburg, A. (2010). The Pencil Code: A High-Order MPI code for MHD Turbulence.

[21] Kosambi, D. D. (1943). Statistics in function space. *Journal of the Indian Mathematical Society*, 7, 76-88.

[22] J. L. Lumley, The structure of inhomogeneous turbulent flows, *Atmospheric Turbulence and Radio Wave Propagation*, edited by A. M. Yaglom and V. I. Tatarsky (Publishing House Nauka, Moscow, USSR, 1967), pp. 166–176.

[23] Sirovich, L. (1987). Turbulence and the dynamics of coherent structures. I. Coherent structures. *Quarterly of applied mathematics*, *45*(3), 561-571.

[24] Meyer, K. E., Pedersen, J. M., & Özcan, O. (2007). A turbulent jet in crossflow analysed with proper orthogonal decomposition. *Journal of Fluid Mechanics*, *583*, 199-227.

[25] Arya, N., Soni, R. K., & De, A. (2015). Investigation of flow characteristics in supersonic cavity using LES. *American Journal of Fluid Dynamics*, *5*(3A), 24-32.

[26] Das, P., & De, A. (2015). Numerical investigation of flow structures around a cylindrical afterbody under supersonic condition. *Aerospace Science and Technology*, *47*, 195-209.66